%% file: final_20_03_2019_arxiv.tex
\documentclass[11pt]{article}

\usepackage[english]{babel}
\usepackage{latexsym,amsmath,amssymb,amsbsy,amstext,amscd,amsfonts,amsthm}
\usepackage{graphics,graphicx}
\usepackage{color}
\usepackage[dvipsnames]{xcolor}
\usepackage{tabularx}
\usepackage{multirow}
\usepackage{booktabs}
\usepackage{float}
\theoremstyle{definition}

\usepackage{array}
\input{definitions}

\date{\today}

\title{Redirection of a crack driven by viscous fluid taking into account plastic effects in the process zone}

\author{M. Wrobel$^{(1)}$, A. Piccolroaz$^{(2)}$, P. Papanastasiou$^{(1)}$ and G. Mishuris$^{(3)}$
\\
{\it $^{(1)}$\! Department of Civil and Environmental Engineering, University of Cyprus, }
\\ {\it 75 Kallipoleos Street, 1678 Nicosia, Cyprus}
\\
{\it $^{(2)}$Dipartimento di Ingegneria Civile, Ambientale e Meccanica, Universita di Trento,}
\\ {\it via Mesiano, 77 I-38123 Trento, Italia}
\\
{\it $^{(3)}$Department of Mathematics, Aberystwyth University,}
\\ {\it Ceredigion SY23 3BZ, Wales, UK}
}

\begin{document}

\maketitle

\begin{abstract}
In this paper the problem of redirection of a crack driven by viscous fluid under  mixed mode loading is considered.The loading includes the classical Modes I - III and the hydraulically induced tangential traction on the fracture walls. The effect of the plastic deformation of the near tip zone is accounted for. Different criteria to determine the fracture deflection angle are examined and compared. 
\end{abstract}

\section{Introduction}

Although the hydraulic fracturing (HF) technique was developed 70 years ago for enhancing the recovery of hydrocarbons from low permeability reservoirs, it became widely known this century as the fracking technology for the exploitation of shale gas and oil resources, with huge success in the United States. Another application of HF, relevant to this study, is in weak and unconsolidated rock formations for combined production stimulation and sand control. This method is known as 'frac-pack' wellbore completion. In all applications, HF involves pumping of a viscous fluid at high rates from a well into the rock formation under high pressure sufficient to fracture the reservoir. The initiated fracture propagates in a complex stress field near the wellbore and re-orients itself to extend further in the direction of the least resistance which is always perpendicular to the minimum in situ compressive stress. During the pumping process, a sand-like material called 'proppant' is mixed with the fracturing fluid. The proppant prevents the fracture from closing after the fluid injection is stopped. Hence, a permeable channel of high conductivity is formed for oil or gas to flow from the reservoir to the well \cite{Economides_2000}. The 'frac-pack' wellbore completion involves additionally, after fracturing the formation, a screen placement and gravel packing of the annulus between screen and rock face for sand control.

Most of the models used by the petroleum industry to simulate and design hydraulic fracturing assume linear-elastic mechanism of the rock deformation and the classic fracture mechanics. The near tip processes include viscous flow in the fracture and formation of a dry zone (fluid-lag), cohesive zone ahead of the crack tip and the surrounding area that is dominated by shear plastic deformation, labeled here a 'process zone'. In weakly consolidated formations growth of short hydraulic fractures is very much determined by the near-fracture-tip plastic deformation of rock coupled with the interaction between the fracturing fluid and fractured layer. Thus, the required propagation pressure and geometry of created fracture depend critically on the tip screening mechanism caused by the plastic rock deformation. An attempt to increase the reliability of numerical simulations of HF and explain the discrepancies between model predictions and field measurements was made by \cite{Papanastasiou_1993,Papanastasiou_1997,van_Dam_2002} who investigated numerically with a fully coupled elastoplastic FEM  HF models the effect of non-linear rock behaviour on the pressure needed for propagation and on the dimensions of the created fractures. Sarris and Papanastasiou \cite{Sarris_2011,Sarris_2013} and Wang \cite{Wang_2015} included the pressure diffusion and porous behavior of the rock in the framework of the cohesive zone model and plastic deformation. All these numerical studies found that plastic yielding provides a shielding mechanism near the tip resulting in an increase of the effective fracture toughness and the resistance to fracturing \cite{Papanastasiou_1999}. Higher pressure is needed to propagate an elastoplastic fracture that appears to be shorter and wider than an elastic fracture. These studies made also clear that the near tip mechanisms are coupled and interact with each other, thus eliminating the relative importance of one mechanism, or may result in an elevated effective fracture toughness at the macroscopic level.

One of the mechanisms that so far has not been studied is the influence of shear stresses that are transferred by the viscous fluid on the crack surfaces near the tip.  Wrobel et al. \cite{wrobel_2017,wrobel_2018} showed that, due to the order of the tip singularity of the hydraulic shear stress in elastic material, this component of the load cannot be omitted when computing the Energy Release Rate (ERR). They introduced a new parameter, called a hydraulic shear stress intensity factor ($K_f$), and proved that it plays an important role in the HF process. In a recent study Papanastasiou and Durban \cite{Papanastasiou_2017} studied the near tip singular plastic fields in a Drucker-Prager power law material and found that the shear stress does not influence the level of singularity but it changes the shape of the developed plastic zones with the emergence of a boundary layer near the fracture surface. Thus, the shear loading and the emergence of a boundary plastic layer influences the 'roughness' of the fracture surface which is very important for the hydraulic conductivity of the fracture \cite{van_Dam_1999}.

In all the studies mentioned above it was assumed that the fracture was initiated and propagated to the preferential direction, perpendicular to the minimum in situ stress. In practice, however, a hydraulic fracture initiates from perforations that are created with shape charges that penetrate the steel casing, cement and near wellbore rock, approximately 30 cm deep. The directions of perforations are not necessarily aligned with the fracture preferential direction as the stress field is unknown and the perforation gun is not usually oriented. Hence the fracture will initiate from perforations and reorient itself gradually as propagating to get aligned far away from the wellbore to the preferential direction.

In this study we will examine the influence of plasticity in the re-orientation of the hydraulic fracture within an analytical solution framework, considering the role of viscous shear stresses. In a recent study \cite{Perkowska_2017} the authors considered the effect of the shear stress induced by the fluid on the crack surfaces to determine the direction of the HF crack propagation in elastic rocks and showed that it may play an essential role in the case of the mixed mode when the total contribution of the classical Stress Intensity Factors (SIF) leads to ERR value close to its critical one. They used the two most popular criteria: the maximum circumferential stress (MCS) criterion \cite{erdogan_1963} and the minimum strain energy density (MSED) criterion \cite{sih_1974}.
It was shown that hydraulic fracture has its own features when the shear traction induced by the fluid on the crack surface originates the local shear stress intensity factor $K_f$ and this leads to significant difference in terms of the crack redirection if the propagation regime is close to viscosity dominated one.
The analysis was based on the recent paper by Wrobel et al. \cite{wrobel_2017} where the authors introduced a new parameter, $K_f$, called the shear stress intensity factor, which plays an important role in accuracy and efficiency of numerical computations. In particular, it was shown that the tip singularity of hydraulically induced shear traction on the crack surfaces is of  order $r^{-1/2}$ as $r \to 0$ and thus has to be taken into account when ERR is computed.

However, it is well known that \emph{inelastic rock deformation near the crack tip} may play an important role in hydraulic fracturing \cite{Papanastasiou_1993,Papanastasiou_1997,Papanastasiou_1999,Papanastasiou_2017}. Thus, the analysys of a combined effect of the plastic deformation and the mixed mode loading (including the shear traction induced by the fluid) on the re-orientation of the fracture is important. There are two ways to tackle the problem of plastic  deformation in the near-tip zone: a) to solve a complete elastic-plastic problem or 
b) to implement a simplified analysis that assumes that the plastic region is much smaller than the zone in which the elastic solution with a dominant singular term holds (i.e. the solution can be fully represented by this singular term). In the latter case the external boundary of the plastic zone is determined from a respective yield criterion.  The first approach, although naturally much more adequate, is simultaneously extremely difficult  to conduct from the technical point of view (see \cite{Bigoni_1993} and references therein). In \cite{Bigoni_1993} the authors considered a Mode I problem with the Drucker-Prager yield criterion. They mentioned that implementation of similar analysis for the mixed mode loading is rather impossible due to its complexity (to the best of our knowledge, no analytical solution has been delivered so far  for the full elastic-plastic problem for a crack under mixed mode loading). Clearly, one can find many numerical research in the area \cite{Camas_2011,Hallback_1994} but their application to parametric analysis is questionable.

Thus, following e.g. \cite{theocaris_T_1982,yehia_Y_1991}, we implement the simplified approach assuming that the plastic deformation zone is small enough in the aforementioned sense and it is enough to consider only the `external' elastic problem, combined with pertinent yield criterion to find the boundary of the plastic zone.

We start from the maximum dilatational strain energy density (MDSED) criterion introduced in \cite{theocaris_T_1982,yehia_Y_1991} to show that, in contrast to the classic fracture mechanics, this criterion in the case of hydraulic fracture does not always allow for predicting the direction of the crack propagation due to the problems with uniqueness and stability of solution. In particular, MDSED criterion fails to produce reasonable results in the  proximity of the so-called viscosity dominated regime \cite{Perkowska_2017}. To eliminate this drawback, we propose a concept of a modified maximum circumferential stress criterion (MMCS) that accounts for local plastic effects by implementing, similarly to MDSED, the plastic zone described by the respective yield condition (von Mises, Drucker-Prager, Tresca or Mohr-Coulomb).
It has been recognized that the applicability of the known criteria is not well justified when accounting for the impact of severe Mode III \cite{lazarus_2008}. Recently, an attempt has been made to tackle such case \cite{cherny_2016,cherny_2017}.
Note that the  MMCS introduced in this paper, provides meaningful results even for a mixed mode with a severe Mode III component.
It is clear that the new criterion should be validated by experiments.

The structure of the paper is as follows. In Section \ref{sec:norm} we introduce necessary notations with reference to \cite{Perkowska_2017} where all details can be found.
 In Section \ref{sec:mdsed} we
consider the classic MDSED criterion. Then, in Section \ref{sec:mmcs}, we introduce a new criterion for the direction of crack propagation: the Modified Maximum Circumferential Stress (MMCS) criterion. Four variants of the criterion are analyzed, each of them utilizing different yield condition).
In Subsection \ref{sec:comparison} all analysed criteria are compared and the impact of the shear stress on the crack propagation direction is discussed. Finally, the conclusions are summed up in Section \ref{sec:conclusions}.

\section{Preliminary results}
\label{sec:norm}

As discussed in \cite{Perkowska_2017}, the stresses around the crack tip in elastic region are distributed in the following manner:
\begin{equation}
\label{eq:asyms}
{\boldsymbol \sigma}(r,\theta,z)=\frac{1}{\sqrt{2\pi r}}\left[ K_I{\bf \Psi}_{I}(\theta)+K_{II}{\bf\Psi}_{II}(\theta)+K_{III}{\bf\Psi}_{III}(\theta)+K_f{\bf\Psi}_{\tau}(\theta)\right]+O\left(\log r\right),
\end{equation}
where $\{r,\theta,z\}$ is a local polar coordinate system traditionally associated with the crack tip,
$K_I$, $K_{II}$ and $K_{III}$ are the classical stress intensity factors (SIFs) and $K_f$ is the shear stress intensity factor related to the hydraulic shear traction acting on the crack surfaces. The functions ${\bf \Psi}_j(\theta)$ define the polar angle dependence and are given by the formulae:

\[
\Psi_{I}^{rr}(\theta) = \frac{1}{4}\left[5\cos \frac{\theta }{2} - \cos \frac{3 \theta}{2}\right],\quad
\Psi_{I}^{\theta\theta}(\theta) = \cos^3\frac{\theta}{2},
\]
\[
\Psi_{I}^{r\theta}(\theta) =
\frac{1}{2}\cos\frac{\theta}{2}\sin\theta,\quad
\Psi_{II}^{rr}(\theta) =
-\frac{1}{4}\left[5\sin \frac{\theta }{2} -3 \sin \frac{3 \theta}{2}\right],
\]
\[
\Psi_{II}^{\theta\theta}(\theta) =-3\sin\frac{\theta}{2}\cos^2\frac{\theta}{2},\quad
\Psi_{II}^{r\theta }(\theta) =\frac{1}{4}\left[\cos\frac{\theta}{2}+3\cos\frac{3\theta}{2}\right],
\]
\[
\Psi_{III}^{r z }(\theta) =\sin\frac{\theta}{2},\quad
\Psi_{III}^{\theta z}(\theta) =\cos\frac{\theta}{2},\quad
\Psi_{\tau}^{rr}(\theta) =-\Psi_{\tau}^{\theta\theta}(\theta)=-\cos \frac{3 \theta}{2},\quad
\Psi_{\tau}^{r\theta}(\theta) =
\sin\frac{3\theta}{2}.
\]
For the plane strain $\Psi^{zz} =\nu \left(\Psi^{rr}+\Psi^{\theta\theta} \right)$.

To define the critical fracture state, we use the Energy Release Rate (ERR) criterion as evaluated in \cite{Perkowska_2017} that accounts for the shear traction induced by the fluid on the crack surfaces:
\begin{equation}
\label{eq:ERR1}
K_I^2 + K_{II}^2  + 4(1-\nu) K_IK_f+\frac{1}{1-\nu} K_{III}^2={\cal E}_C\equiv K^2_{IC},
\end{equation}
where $\nu$ is the Poisson's ratio, $E$ is the Young's modulus, while ${\cal E}_C$ and $K_{IC}$ are the critical values determining ERR and material toughness, respectively, and need to be found experimentally.

To decrease the number of unknown parameters during the parametric study below, the problem is normalised by introducing the following natural scaling:
\begin{equation}
\label{eq:Knorm}
\hat K_I =\frac{K_I}{K_{IC}},\quad \hat K_{II} =\frac{K_{II}}{K_{IC}},\quad \hat K_{III} =\frac{K_{III}}{K_{IC}},\quad \hat K_f=\frac{K_f}{K_{IC}}.
\end{equation}
The fracture criterion \eqref{eq:ERR1} then becomes:
\begin{equation}
\label{eq:ERRnorm}
\hat K_I^2 + \hat K_{II}^2 + 4(1-\nu) \hat K_I \hat K_f + \frac{1}{1-\nu} \hat K_{III}^2 =1.
\end{equation}
Note that in such formulation the value of the material toughness is completely hidden and the large and small toughness regimes cannot be recognised or identified immediately.
To make this possible, we introduce a dimensionless parameter $\tilde p_0=2\pi p_0(1-\nu^2)/E $, where $p_0$ is a multiplier of the leading (logarithmic) term in the asymptotic expansion of the fluid pressure near the crack tip (see \cite{wrobel_2017} or \cite{wrobel_2015}). It can be used to produce the following interrelation between $\hat K_I$ and $\hat K_f$  (compare with equation $(71)$ from \cite{wrobel_2017}):
\begin{equation}
\label{eq:Kfnorm}
 \varpi=\frac{\tilde p_0}{\pi(1-\nu)-\tilde p_0}, \quad \hat K_f=\varpi \hat K_I,\quad 0<\tilde p_0<\pi(1-\nu).
\end{equation}

The values of the parameter $\tilde p_0$ and the stress intensity factors are not independent.
As it was shown in \cite{wrobel_2017} for the Mode I deformation ($K_{II}=K_{III}=0$), the value of the parameter $\tilde p_0$ determines
the propagation regime ($\tilde p_0\to0$ corresponds to toughness dominated one while $\tilde p_0\to \pi(1-\nu)$ determines the viscosity dominated regime).

Following \cite{Perkowska_2017} we conclude that:
\begin{equation}
\label{nob_mode1_p0}
K_{IC}\cdot \hat K^{\text{eff}}_{IC}\to\infty \quad  \Leftrightarrow  \quad
\hat K_I \to 1\quad \mbox{and}\quad  \tilde p_0 \to 0,
\end{equation}
and
\begin{equation}
\label{nob_mode1_p01}
K_{IC}\cdot \hat K^{\,e\!f\!f}_{IC}\to0 \quad  \Leftrightarrow \quad
\hat K_I \to 0  \quad \text{and} \quad \tilde p_0 \to \pi(1-\nu),
\end{equation}
where the normalised effective toughness is defined as follows:
\begin{equation}
\label{nob_mode1_p01}
\hat K^{\,e\!f\!f}_{IC}=\sqrt{1-\hat K_{II}^2-\frac{1}{1-\nu}\hat K_{III}^2}.
\end{equation}

Combining  \eqref{eq:ERRnorm} and \eqref{eq:Kfnorm}  provides, after rearrangement, a formula for $\hat K_I$ when $\hat K_{II}$,  $\hat K_{III}$ and $\tilde p_0$ are known:
\begin{equation}
\label{eq:KInorm}
\hat K_I=\sqrt{\frac{\pi(1-\nu)-\tilde p_0}{\pi(1-\nu)+\tilde p_0(3-4\nu)}}\,\hat K^{\,e\!f\!f}_{IC}.
\end{equation}

After substitution of \eqref{eq:Kfnorm} into \eqref{eq:KInorm} and some algebra we obtain a relation for $\hat K_f$:
\begin{equation}
\label{eq:Kform2}
\hat K_f=\frac{\tilde p_0}{\sqrt{\big[\pi(1-\nu)-\tilde p_0\big]\big[\pi(1-\nu)+\tilde p_0(3-4\nu)\big]} }\,\hat  K^{\,e\!f\!f}_{IC}.
\end{equation}
It can be easily seen that for any fixed values of $\hat K_{II}$ and $\hat K_{III}$ one has (compare \eqref{eq:KInorm} and \eqref{eq:Kform2}):
\begin{equation}
\label{eq:limK}
\lim_{\tilde p_0\to\pi(1-\nu)}\hat K_f\hat K_I=\frac{1}{4(1-\nu)}\Big(\hat K^{\,e\!f\!f}_{IC}\big)^2,
\end{equation}
Finally, note that for $\hat K_{III}=0$ and $\tilde p_0=0$ (classic mixed Mode I and II), both normalised stress intensity factors, $\hat K_I=\sqrt{1-\hat K_{II}^2}$ and $\hat K_f=0$, are independent of $\nu$.

Equations \eqref{eq:KInorm} and \eqref{eq:Kform2} provide a relationship between the normalised symmetric SIFs, $\hat K_I$ and $\hat K_f$, and the normalised anti-symmetric SIFs, $\hat K_{II}$ and $\hat K_{III}$, taking into account also the influence of the hydraulically induced shear stresses through the pressure parameter $\tilde{p}_0$. This allows for a parametric study of the fracture propagation angle, where the independent parameters are $\hat K_{II}$, $\hat K_{III}$ and $\tilde{p}_0$. This analysis is given in the next sections.

\section{Maximum Dilatational Strain Energy Density (MDSED) criterion}
\label{sec:mdsed}

Let us begin by analysing the maximum dilatational strain energy density  (MDSED) criterion, proposed by Theocaris and Andrianopoulos in \cite{theocaris_T_1982}. Noting that the total strain energy density, $W$, can be divided into two components at any given point -- the dilatational (volumetric) strain energy density, $W_v$, and the distortional strain energy density, $W_d$ -- we have that:
\begin{equation}\label{dmsed_2comp}
W=W_v+W_d,
\end{equation}
for:
\begin{equation}
\label{Wv}
W_v=\frac{1}{18 K} (\tr\bsigma)^2=\frac{1 - 2\nu}{6 E}\left[ \sigma_{rr}+\sigma_{\theta \theta}+ \sigma_{zz}\right]^2,
\end{equation}
\begin{multline}
\label{Wd}
W_d=\frac{1}{4G} \dev\bsigma\cdot\dev\bsigma= \\[1mm]
\frac{1 + \nu}{3 E} \left(\sigma_{rr}^2+\sigma_{\theta \theta}^2+\sigma_{zz}^2-\sigma_{rr}\sigma_{\theta \theta}-\sigma_{\theta \theta}\sigma_{zz}-\sigma_{zz}\sigma_{rr}+3(\sigma_{r\theta}^2+\sigma_{rz}^2+\sigma_{\theta z}^2)\right),
\end{multline}
where $\sigma_{zz}=\nu\left( \sigma_{rr}+\sigma_{\theta\theta} \right)$ for plane strain and $K=E/[3(1-2\nu)]$ is the volumetric modulus of elasticity (bulk modulus).

The criterion assumes that the radius of the elastic-plastic boundary, $r_b$, is determined from the von Mises yield condition to produce a surface of constant distortional strain energy density. Then a maximum of the dilatational strain energy density is searched along this surface. This maximum determines the angle of crack propagation.

According to the von Mises yield condition, the critical value of the distortional strain energy density, $W_d$, obtained for the elastic-plastic boundary, is: 
%
\begin{equation}\label{vm}
W_d =  \frac{1+\nu}{3E}\sigma_t^2,
\end{equation}
with $\sigma_t$ being the uniaxial yield strength.

The corresponding radius of the plastic deformation zone, $r_b$, was derived as: 

\begin{equation}
\label{rb_vm}
r_b=R_b{\tilde r}_b(\theta), \quad \theta\in(-\pi,\pi),
\end{equation}
where
\begin{equation}
\label{R_b}
R_b=\frac{K_{IC}^2}{\sigma_t^2}, 
\end{equation}

has a dimension of length [m] and the dimensionless part

\begin{equation}
\label{rb_vm_theta}
\tilde{r}_b=\frac{1}{8\pi}\left\{\hat K_I^2 c_1(\theta)+ \hat K_{II}^2 c_2(\theta)+ \hat K_I \hat K_{II} c_3(\theta)+
\hat K_f^2 c_4(\theta)+  \hat K_I \hat K_f c_5(\theta)+ \hat K_{II} \hat K_f c_6(\theta)+12\hat K_{III}^2\right\},
\end{equation}
describes the shape of the zone and simultaneously affects its size  by the values of the parameters (SIFs and the Poison's ratio $\nu$).

Here by $c_j(\theta)$ we have defined the following dimensionless trigonometrical factors:
\begin{align}
\label{constants_c}
c_1(\theta)&=2 \cos^2\frac{\theta}{2} \left(5 - 8 \nu\left(1-\nu\right) - 3 \cos\theta\right), \quad c_3(\theta)=4 \sin\theta\left(3\cos\theta-\left(1 - 2 \nu\right)^2\right),\nonumber\\
c_2(\theta)&= 5 - 8 \nu\left(1-\nu\right) - 2 \left(1 - 2 \nu\right)^2 \cos\theta + 9 \cos^2 \theta,\\
c_4(\theta)&= 48 \left(1-\nu\right)^2, \quad
c_5(\theta)=24 (1- \nu) \sin^2\theta, \quad
c_6(\theta)=12 (1 -\nu)\sin 2\theta,\nonumber
\end{align}

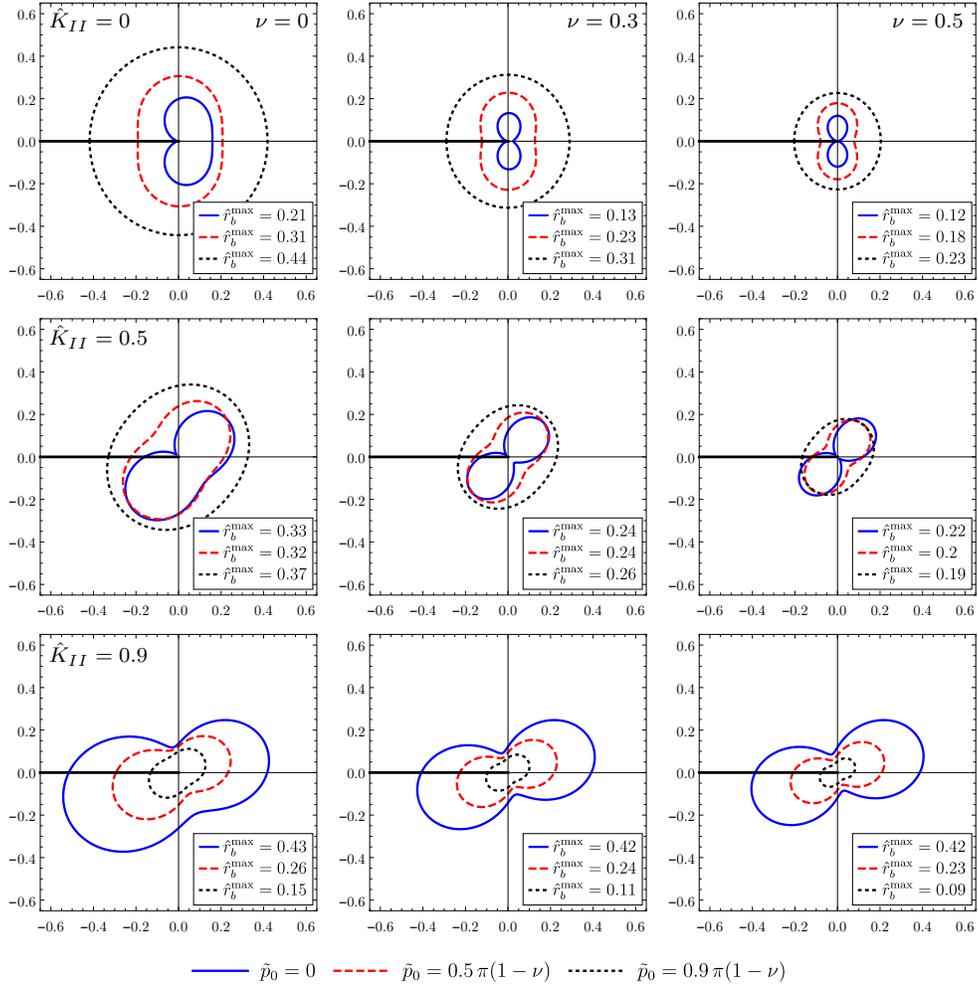
\begin{figure}[htb!]
\begin{center}
\includegraphics[scale=0.75]{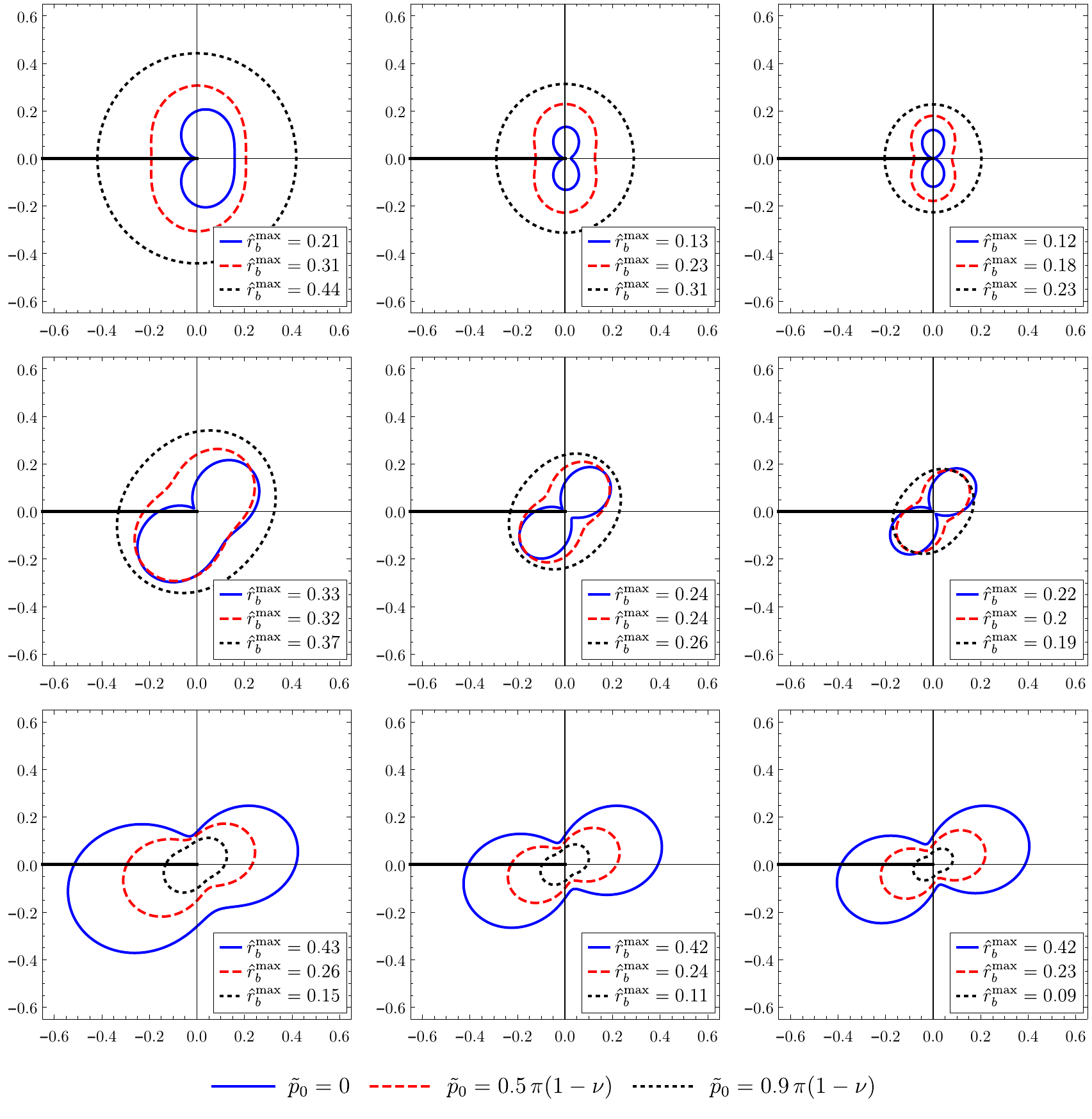}
{
\scriptsize
\put(-352,368){$\hat K_{II}=0$}
\put(-352,248){$\hat K_{II}=0.5$}
\put(-352,128){$\hat K_{II}=0.9$}
\put(-274,368){$\nu=0$}
\put(-156,368){$\nu=0.3$}
\put(-33,368){$\nu=0.5$}
}
\caption{MDSED: The shapes of of the plastic zones described by the normalized radius $\hat r_b$ \eqref{r_hat} for various values of $\tilde p_0$ and fixed $\hat K_{II}$ and $\nu$. The blue lines reflect the actual sizes of the plastic zones, the red ones utilize the scaling factor 0.5, while the black curves are multiplied by 0.1. }
\label{MDSED_rc}
\end{center}
\end{figure}

The relative sizes and shapes of the plastic zones are depicted in Fig.~\ref{MDSED_rc} for three values of $\hat K_{II}=\{0,0.5,0.9\}$ and $\nu=\{0,0.3,0.5\}$ by means of a normalized radius:
\begin{equation}
\label{r_hat}
\hat r_b={\tilde r}_b \frac{\pi(1-\nu)-\tilde p_0}{\pi(1-\nu)}.
\end{equation}

This additional scaling has been introduced to make the curves that refer to different crack propagation regimes distinguishable in one picture. Indeed, under this scaling, the blue lines reflect the actual values of ${\tilde r}_b$, the red ones utilize the scaling factor 0.5, while the black curves are multiplied by 0.1.

Clearly, a symmetry of the plastic zone with respect to the plane of crack propagation (horizontal axis) is obtained only for the symmetrical loading ($\hat K_{II}=0$). When approaching the viscosity dominated regime the shape of the zone becomes elliptic regardless of the magnitude of $\hat K_{II}$ and a  value of the Poisson ratio. Conversely, as moving towards the toughness dominated mode, the plastic zone contour tends to be more irregular, which is especially pronounced for the severe antisymmetric loading ($\hat K_I \ll \hat K_{II}$).

When analyzing the size of plastic yield area we can see two different trends for the limiting regimes. Namely, for the viscosity dominated mode the plastic zone shrinks with growing $\hat K_{II}$, while for the toughness dominated regime a reverse mechanism is observed. Obviously, the character of the zone evolution in the interim is governed by those two counteracting tendencies. It is always the viscosity dominated regime which provides the greatest size of the plastic area, and the toughness dominated mode which results in the smallest extent of it. For all considered cases an increase of $\nu$ reduces the area of plastic zone.

The criterion described in \cite{theocaris_T_1982} states that the maximum value of the dilatational strain energy density, $W_v(r,\theta)$, has to be found along the elastic-plastic boundary.
In other words, one needs to analyse an auxiliary function:
\begin{equation}
\label{F_1_def}
F_1(\theta)=\frac{3E}{8(1+\nu)^2 (1 - 2 \nu)}W_v({\tilde r}_b(\theta),\theta),
\end{equation}

%
where the crack propagation direction, $\theta_f$, is found from the condition:
\begin{equation}
\label{dmsed_sol}
\theta_f=\theta \Big|_{\{F_1=F_1^{\text{max}}\}\wedge \{\sigma_{\theta\theta}>0\}}.
\end{equation}
As it was in the case of the Minimum Strain Energy Density criterion (see \cite{Perkowska_2017}), to preserve the physical sense of solution, the maximum should be sought for under additional condition that the circumferential stress is positive.

According to \cite{Perkowska_2017} the expression for $\sigma_{\theta \theta}$ has the following form:
\begin{equation}
\label{sigma_tt}
\sigma_{\theta \theta}(r,\theta)=\frac{K_{I\text{c}}}{\sqrt{2\pi r}}\left[\hat K_I \cos^3\frac{\theta}{2}-3\hat K_{II}\sin \frac{\theta}{2}\cos^2\frac{\theta}{2}+2(1-\nu)\hat K_\text{f}\cos \frac{3\theta}{2}
 \right].
\end{equation}

\begin{figure}[htb!]
\begin{center}
\includegraphics[scale=0.7]{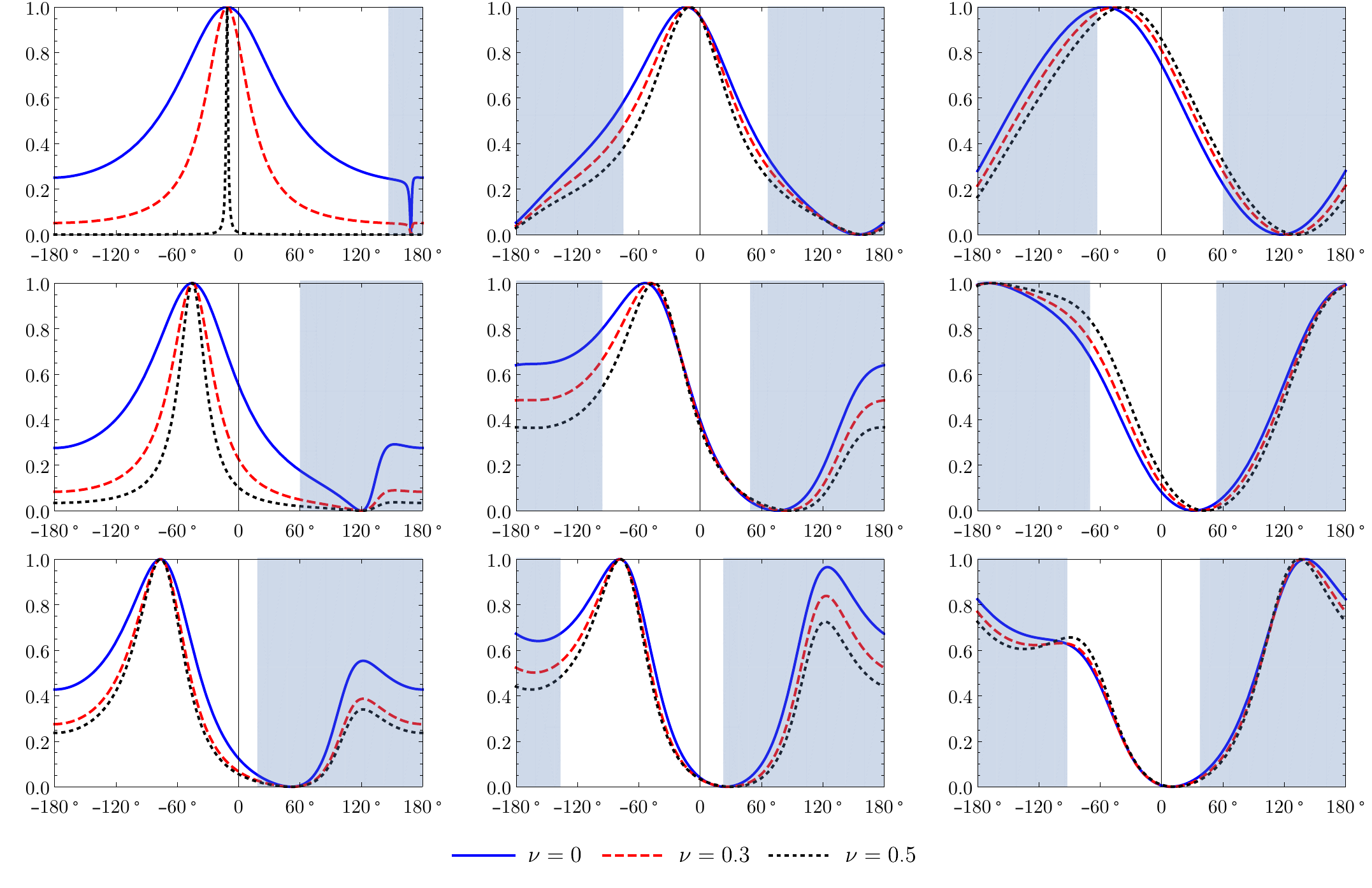}
{
\tiny
\put(-440,62){\rotatebox{90}{$\ds F_1(\theta)/F_1^{max}$}}
\put(-440,150){\rotatebox{90}{$\ds F_1(\theta)/F_1^{max}$}}
\put(-440,238){\rotatebox{90}{$\ds F_1(\theta)/F_1^{max}$}}
\put(-362,15){$\theta$}
\put(-215,15){$\theta$}
\put(-68,15){$\theta$}
\put(-416,271){$\hat K_{II}=0.1$}
\put(-416,184){$\hat K_{II}=0.5$}
\put(-416,35){$\hat K_{II}=0.9$}
\put(-347,269){$\ds \frac{\tilde p_0}{\pi(1-\nu)}=0$}
\put(-206,269){$\ds \frac{\tilde p_0}{\pi(1-\nu)}=0.5$}
\put(-59,269){$\ds \frac{\tilde p_0}{\pi(1-\nu)}=0.9$}
}
\caption{MDSED: Value of $F_1(\theta)/F_1^{max}$ for various values of Poisson's ratio and fixed $\hat K_{II}$ and $\tilde p_0$. The grey regions on the graphs correspond to the areas where $\sigma_{\theta\theta} < 0$.}
\label{MDSED_func}
\end{center}
\end{figure}
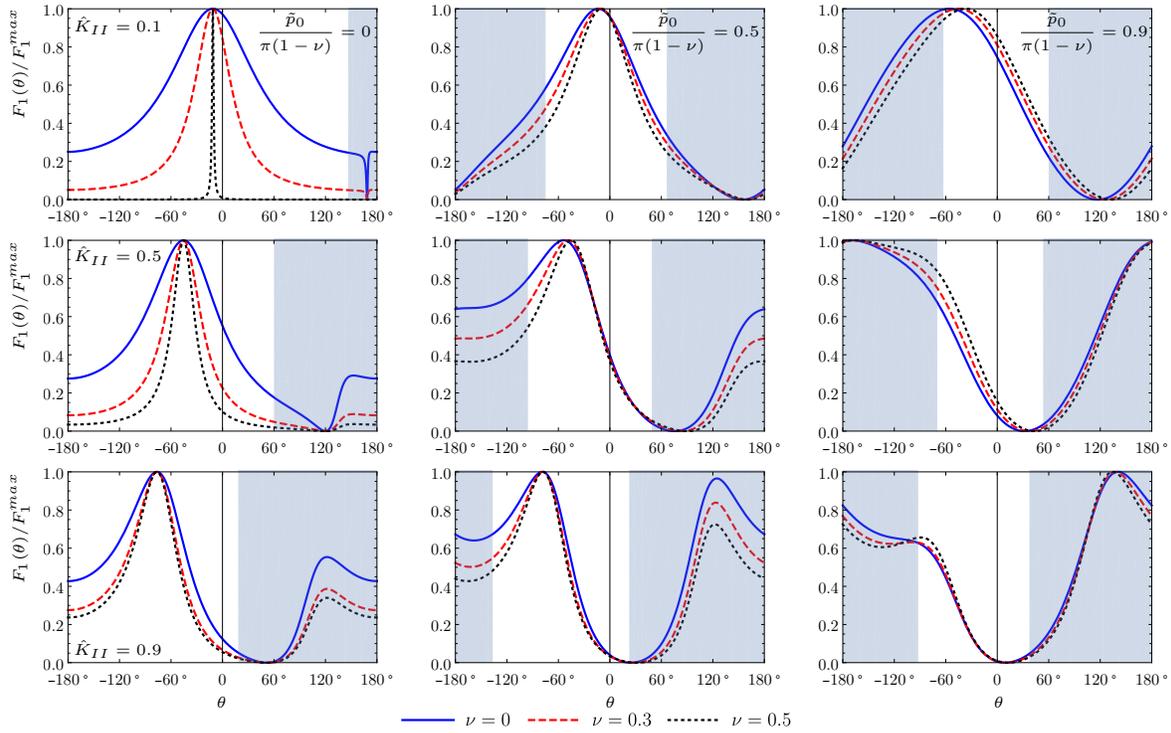

In Fig.~\ref{MDSED_func} values of the function $F_1(\theta)$ normalised by its maximum  are given for $\tilde p_0=0$, $\tilde p_0=0.5\pi(1-\nu)$ and $\tilde p_0=0.9\pi(1-\nu)$. We do not show the results for the viscosity dominated regime, because when approaching $\tilde p_0=\pi(1-\nu)$ there is either no local maximum for $\theta<0$  that would provide continuity of the solution, or there is no value of $\theta_f$ that satisfies the condition of positive circumferential stress. Even for the analyzed liming case $\tilde p_0=0.9\pi(1-\nu)$ one can observe that, depending on the values of $\hat K_{II}$ and $\nu$, there may not exist a local maximum of $F_1$ inside the interval of $\theta$ that  corresponds to  $\sigma_{\theta \theta}(r,\theta)>0$.  For this reason we will consider in the following two variants of the criterion for the case when the condition of local maximum of $F_1$ is not satisfied:
\begin{itemize}
\item{variant I - we assume that there is locally no solution to the problem (or in other words, the criterion does not predict the propagation angle over certain range of the loading parameters);}
\item{variant II - we take the value of $\theta$ which corresponds to the global maximum of  $F_1$ in the domain $\sigma_{\theta \theta} \geq 0$;}
\end{itemize}

\begin{figure}[htb!]
\begin{center}
\includegraphics[scale=0.5]{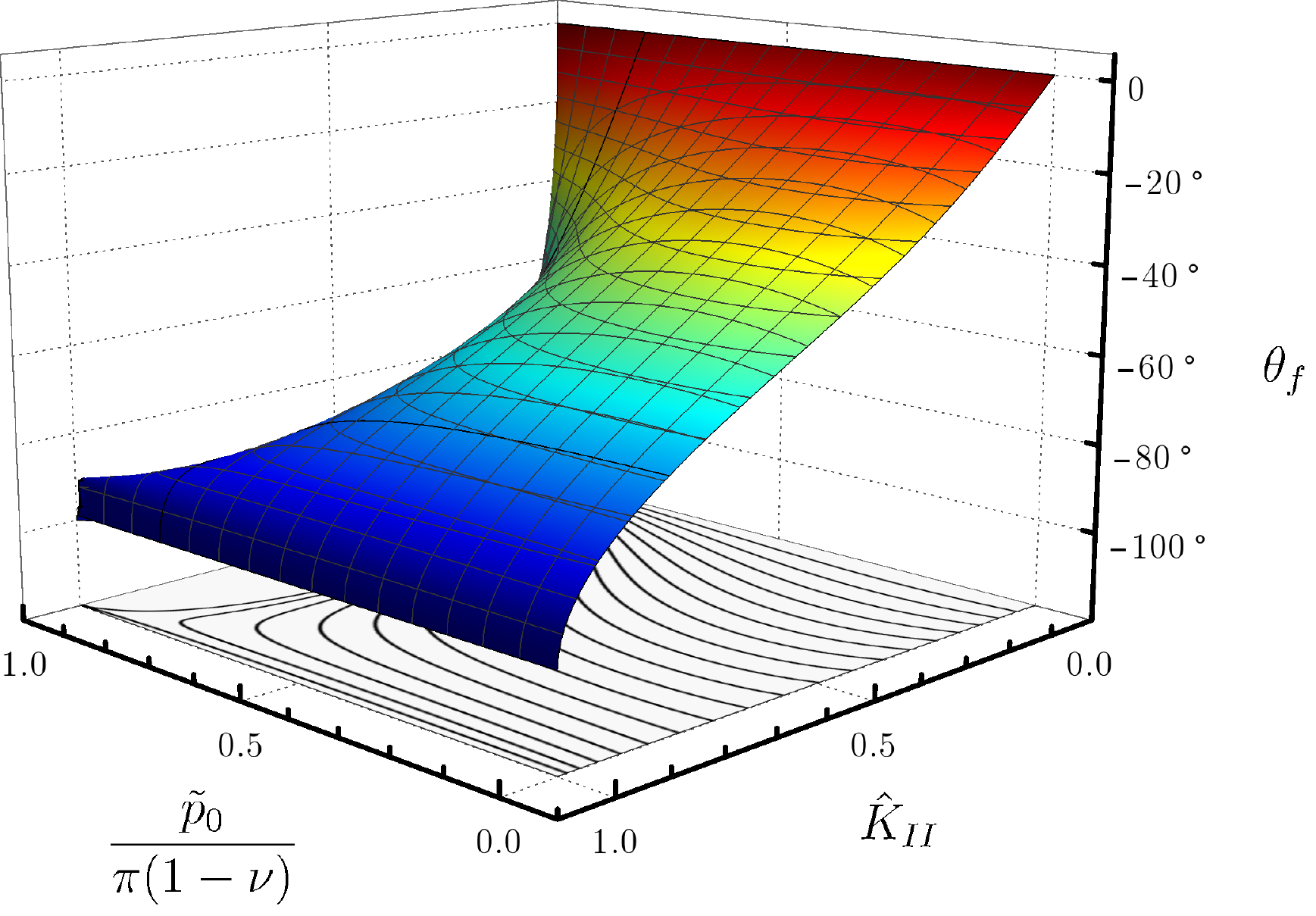}
\hspace{0mm}
\includegraphics[scale=0.45]{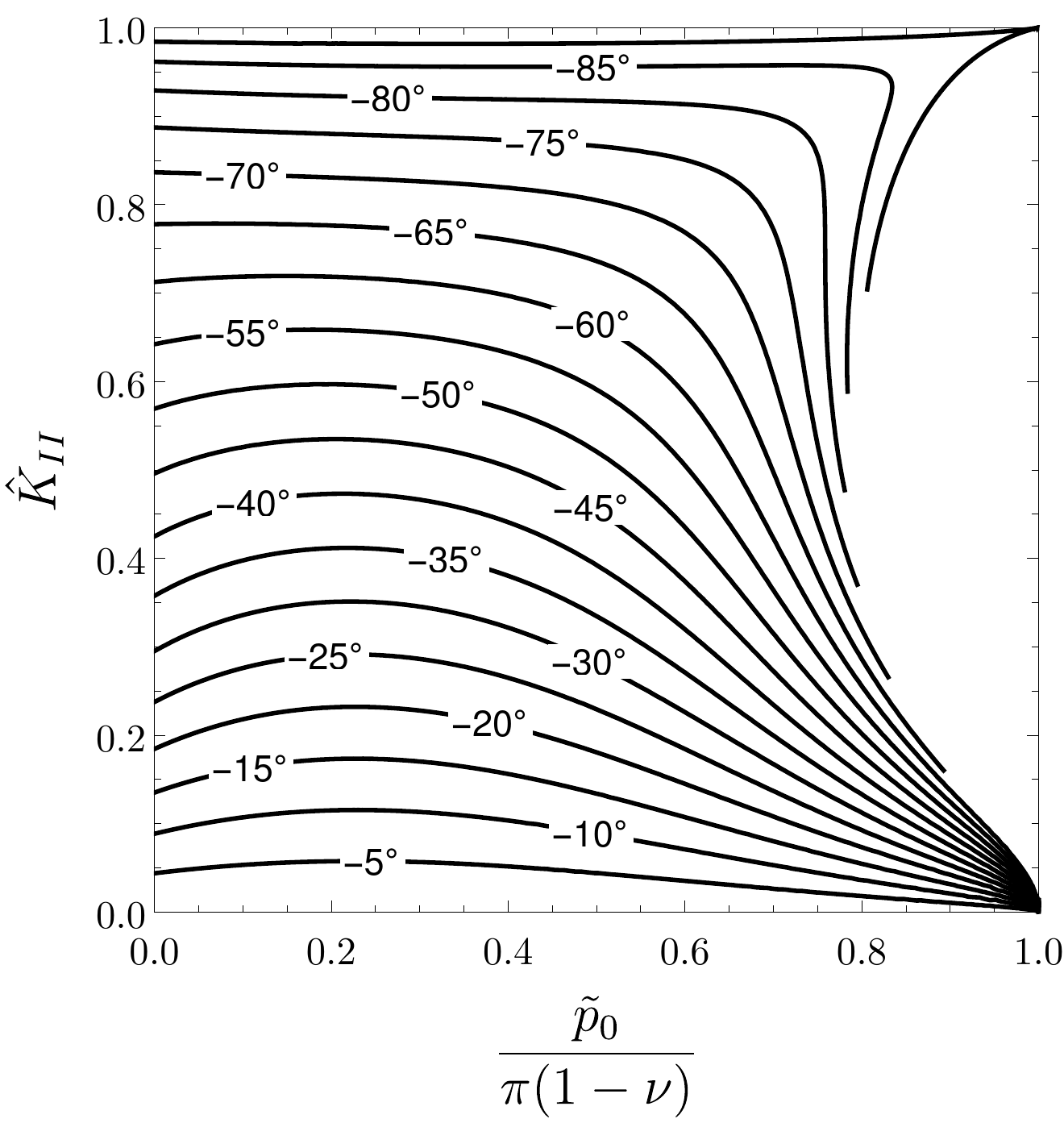}
\put(-165,20){$A$}
\put(5,20){$B$}
\put(-165,180){$D$}
\put(5,180){$C$}
\caption{MDSED: Predicted propagation angle $\theta_f$ for $\hat K_{II} \in [0,1]$ and $\tilde p_0 \in [0,\pi(1-\nu)]$ for $\nu=0.3$. The local maximum of $F_1$ is accepted (variant I).}
\label{MDSED_n03}
\end{center}
\end{figure}

\begin{figure}[htb!]
\begin{center}
\includegraphics[scale=0.5]{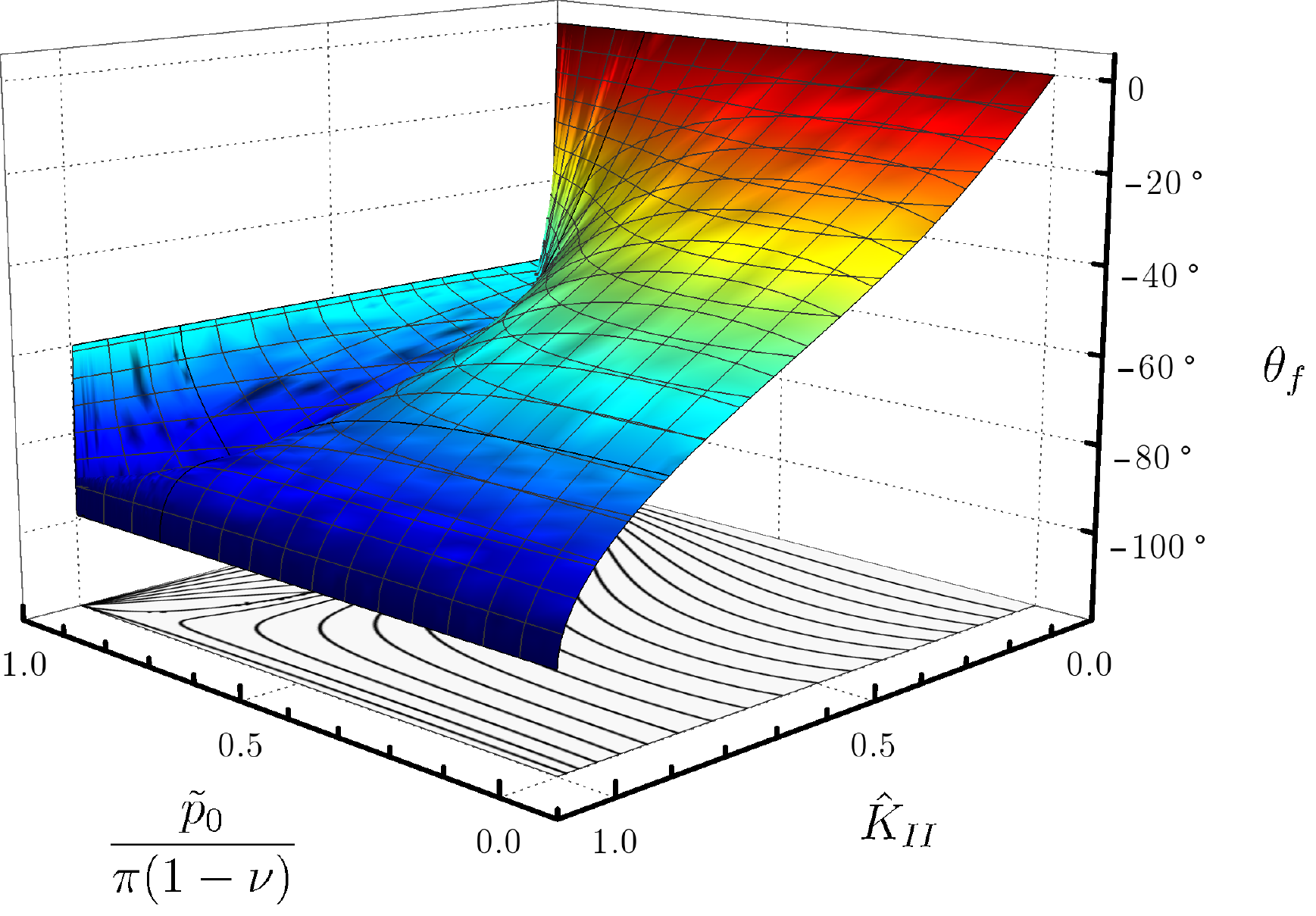}
\hspace{0mm}
\includegraphics[scale=0.45]{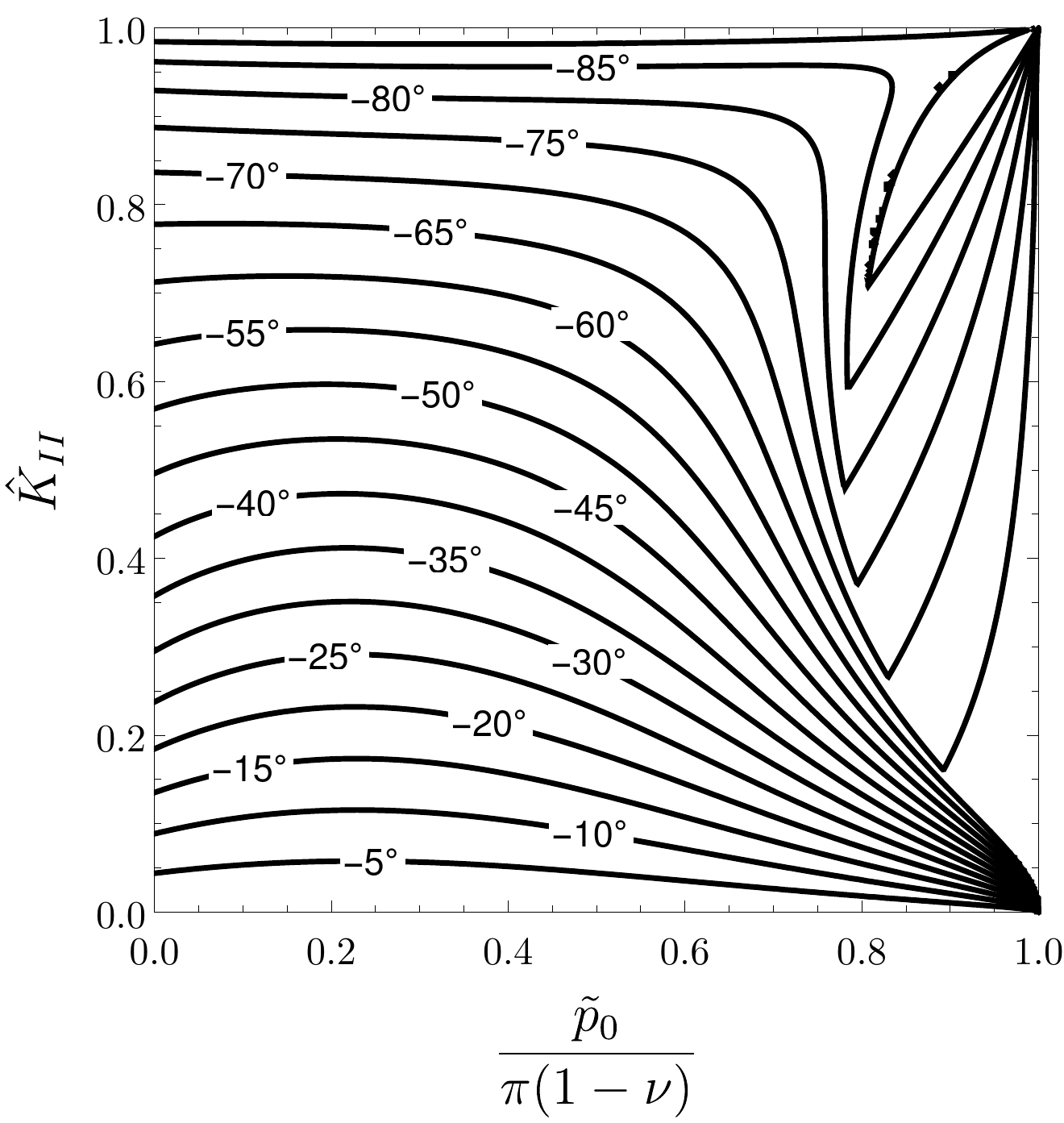}
\put(-165,20){$A$}
\put(5,20){$B$}
\put(-165,180){$D$}
\put(5,180){$C$}
\caption{MDSED: Predicted propagation angle $\theta_f$ for $\hat K_{II} \in [0,1]$ and $\tilde p_0 \in [0,\pi(1-\nu)]$ for $\nu=0.3$. The global maximum of $F_1$ is accepted (variant II).}
\label{MDSED2_n03}
\end{center}
\end{figure}

\begin{figure}[htb!]
\begin{center}
\includegraphics[scale=0.7]{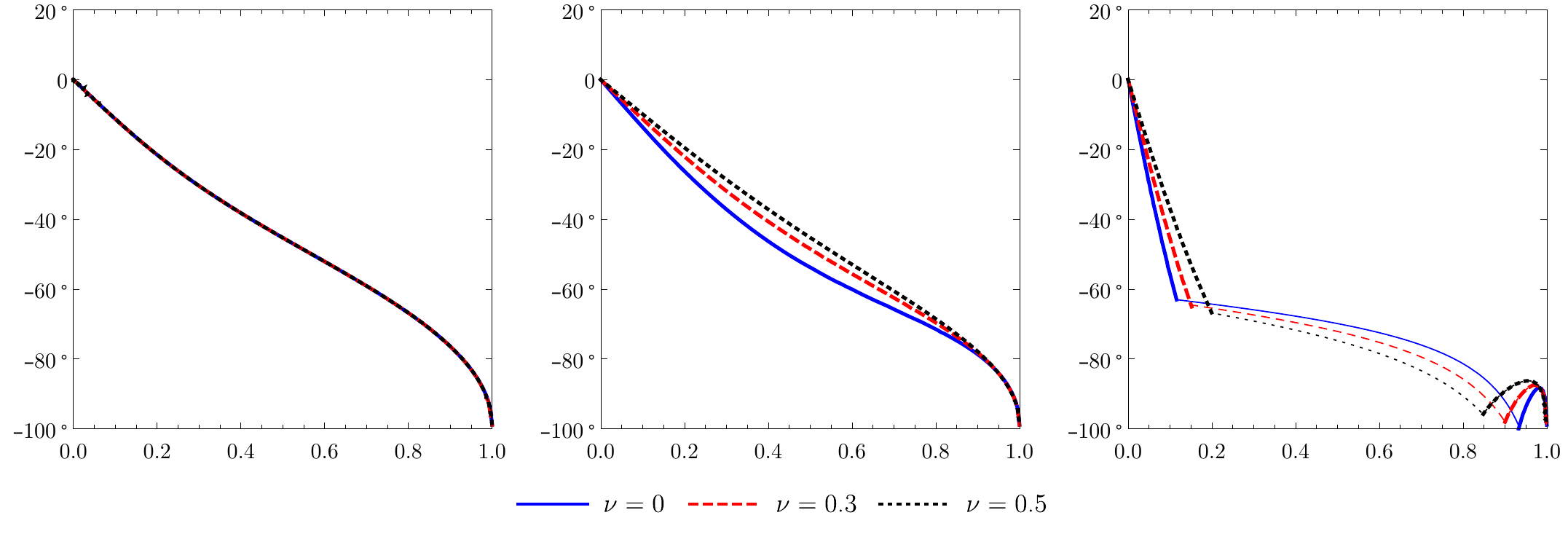}
{
\scriptsize
\put(-433,138){$\theta_f$}
\put(-287,138){$\theta_f$}
\put(-140,138){$\theta_f$}
\put(-360,17){$\hat K_{II}$}
\put(-214,17){$\hat K_{II}$}
\put(-66,17){$\hat K_{II}$}
\put(-350,133){$\ds \frac{\tilde p_0}{\pi(1-\nu)}=0$}
\put(-210,133){$\ds \frac{\tilde p_0}{\pi(1-\nu)}=0.5$}
\put(-64,133){$\ds \frac{\tilde p_0}{\pi(1-\nu)}=0.9$}
}
\caption{MDSED: Value of $\theta_f$ for various values of Poisson's ratio.}
\label{MDSED_nu}
\end{center}
\end{figure}
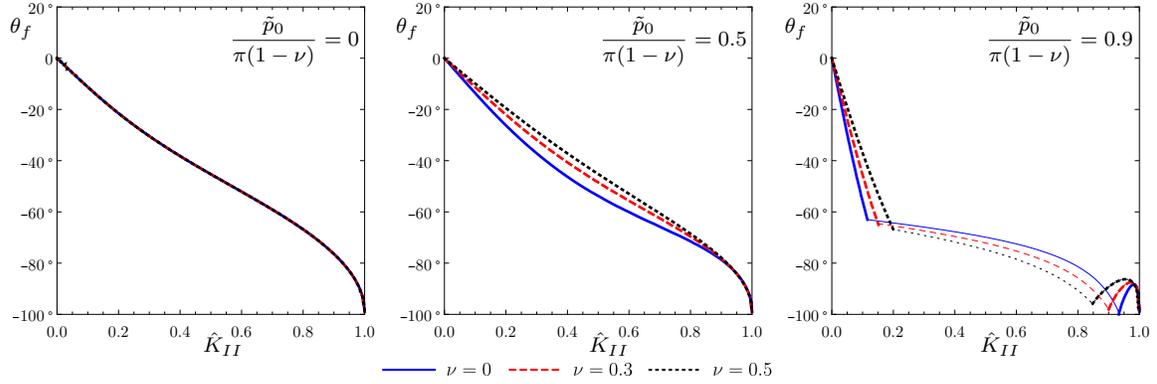

Respective results for $\nu=0.3$ and all admissible values of $\hat K_{II} \in [0,1]$ and $\tilde p_0 \in [0,\pi(1-\nu)]$ are presented in Fig.\ref{MDSED_n03} (variant I) and Fig.\ref{MDSED2_n03} (variant II). In Fig.\ref{MDSED_n03} (variant I)  one can clearly see the region in the vicinity of the viscosity dominated regime (edge BC), over which the solution for $\theta_f$ does not exist. Even if this drawback is eliminated in the variant II (Fig.\ref{MDSED2_n03}), the solution in this area behaves in a rather peculiar way. Pertinent trends can be easily identified in Fig.\ref{MDSED_nu}c). It shows that when employing variant II, the angle of crack propagation is only a continuous (but locally not smooth) function of $\hat K_{II}$ and any small perturbation to the value of $\hat K_{II}$ around the locations of $\theta_f '$ discontinuities in the considered parametric space can result in a serious change of  $\theta_f $. These facts cast doubts on the applicability of the MDSED criterion in the case when hydraulically induced tangential tractions are accounted for.
Note that outside the discussed area both considered variants yield the same results. Clearly, for $\hat K_{II}=0$ (edge AB) one has $\theta_f=0$. Moreover, for $\hat K_{II}=1$ (edge CD) the solution can be found analytically:

\begin{equation}
\theta_f=-\arccos\left(1-\frac{2}{\sqrt{3}}\right).
\end{equation}

Notably, in the corners B and C ($\hat K_{II}=0$ and $\hat K_{II}=1$ for the viscosity dominated mode) $\theta_f$ has no limit, which effectively means that the angle of crack propagation depends here crucially on the load history.


To complement the analysis of the MDSED criterion we show in Fig.\ref{MDSED_nu} the sensitivity of the crack propagation angle to the value of the Poisson's ratio. It turns out that for the toughness dominated regime one obtains the same results regardless of the value of $\nu$ (the same trend was reported in \cite{Perkowska_2017} for the Minimum Strain Energy Density Criterion). Some differences between the results for various $\nu$ can be observed when moving towards the viscosity dominated mode, however the influence of the Poisson's ratio on $\theta_f$ is rather limited. The maximal disparities between the crack propagation angles for set value of $K_{II}$ amount to 10$^\circ$ (that is around 20 $\%$ of the absolute value of $\theta_f$).

\section{Modified Maximum Circumferential Stress (MMCS) criterion}
\label{sec:mmcs}

In order to avoid the mentioned above drawbacks of MDSED criterion we propose here a new one. It is based on the assumption that the direction of fracture propagation is defined by the maximal value of the circumferential stress (compare \cite{Perkowska_2017}) obtained along the elastic-plastic boundary in the elastic domain. In this way the new approach
incorporates the plastic yield stress criterion, depends on the Mode III component and preserves all the advantages of the MCS criterion \cite{Perkowska_2017}. From now on it will be referred to as the Modified Maximum Circumferential Stress (MMCS) criterion.

Below, we analyze four  variants of the criterion, each of them based on a different yield condition i.e.: the von Mises condition, the Drucker-Prager condition, the Tresca condition or the Mohr-Coulomb condition.

\subsection{MMCS criterion: von Mises yield criterion (MMCS-vM)}

Let us first analyse the MMCS under von Mises yield criterion. As it was in the case of the MDSED approach, the elastic-plastic boundary, $r_b$, is found from \eqref{rb_vm}.The shapes and sizes of plastic zones are shown in Fig. \ref{MDSED_rc} with the respective comments included in Subsection \ref{sec:mdsed}.

In order to determine the angle of fracture deflection one needs to find the maximum of the following function:
\begin{equation}
\label{F2_def}
F_2(\theta)=\sigma_{\theta\theta}\left(r_b(\theta),\theta\right),
\end{equation}
%
and hence:
\begin{equation}
\label{mmcsvm_sol}
\theta_f=\theta \Big|_{F_2=F_2^{\text{max}}}.
\end{equation}
where the circumferential stress, $\sigma_{\theta \theta}$ is computed according to \eqref{sigma_tt}.

\begin{figure}[htb!]
\begin{center}
\includegraphics[scale=0.7]{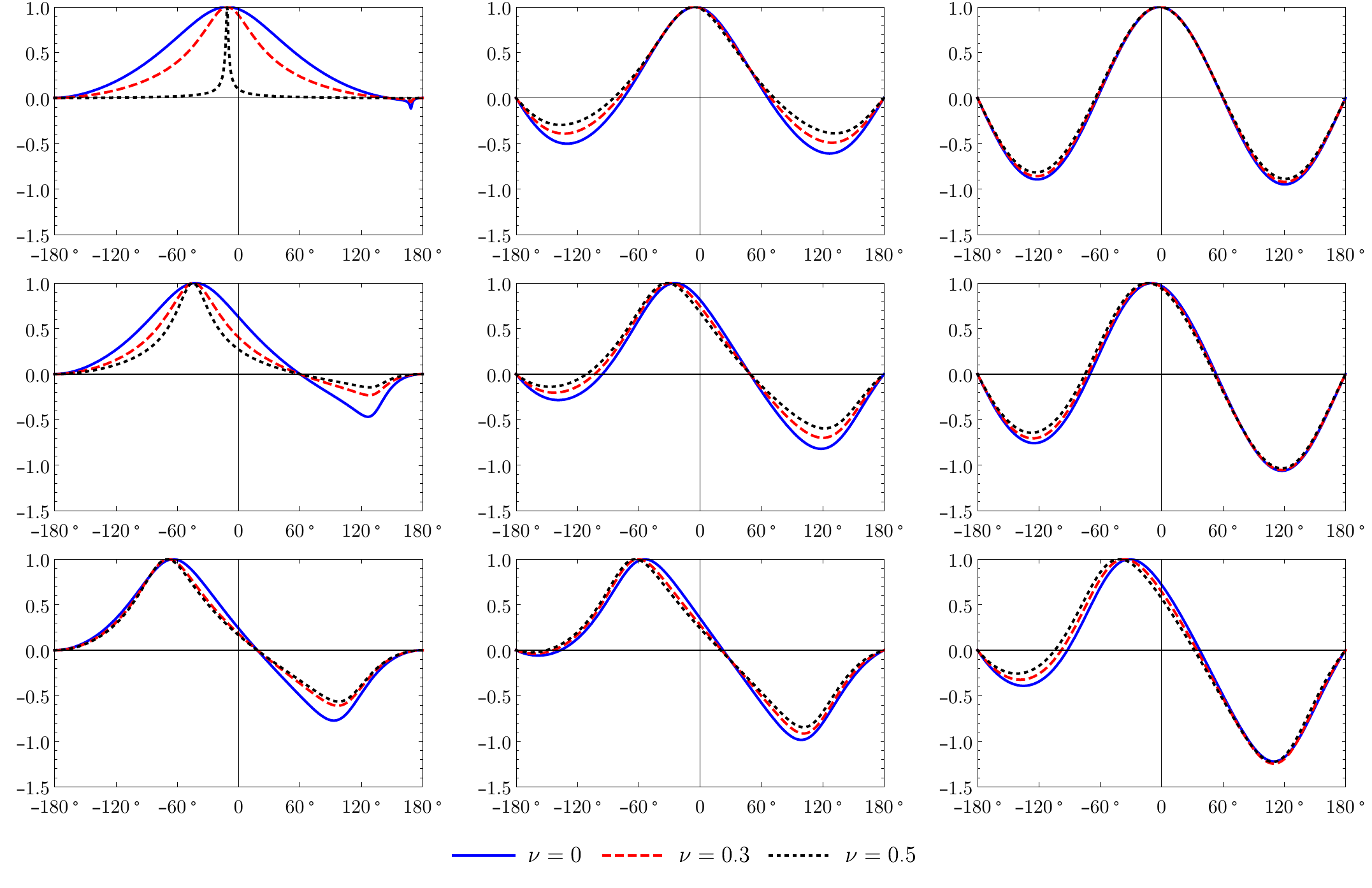}
{
\tiny
\put(-440,62){\rotatebox{90}{$\ds F_2(\theta)/F_2^{max}$}}
\put(-440,150){\rotatebox{90}{$\ds F_2(\theta)/F_2^{max}$}}
\put(-440,238){\rotatebox{90}{$\ds F_2(\theta)/F_2^{max}$}}
\put(-362,15){$\theta$}
\put(-215,15){$\theta$}
\put(-68,15){$\theta$}
\put(-416,215){$\hat K_{II}=0.1$}
\put(-416,125){$\hat K_{II}=0.5$}
\put(-416,38){$\hat K_{II}=0.9$}
\put(-347,215){$\ds \frac{\tilde p_0}{\pi(1-\nu)}=0$}
\put(-206,215){$\ds \frac{\tilde p_0}{\pi(1-\nu)}=0.5$}
\put(-59,215){$\ds \frac{\tilde p_0}{\pi(1-\nu)}=0.9$}
}
\caption{MMCS-vM: Value of $F_2(\theta)/F_2^{max}$ for various values of Poisson's ratio and fixed $\hat K_{II}$ and $\tilde p_0$.}
\label{MMCS_func}
\end{center}
\end{figure}
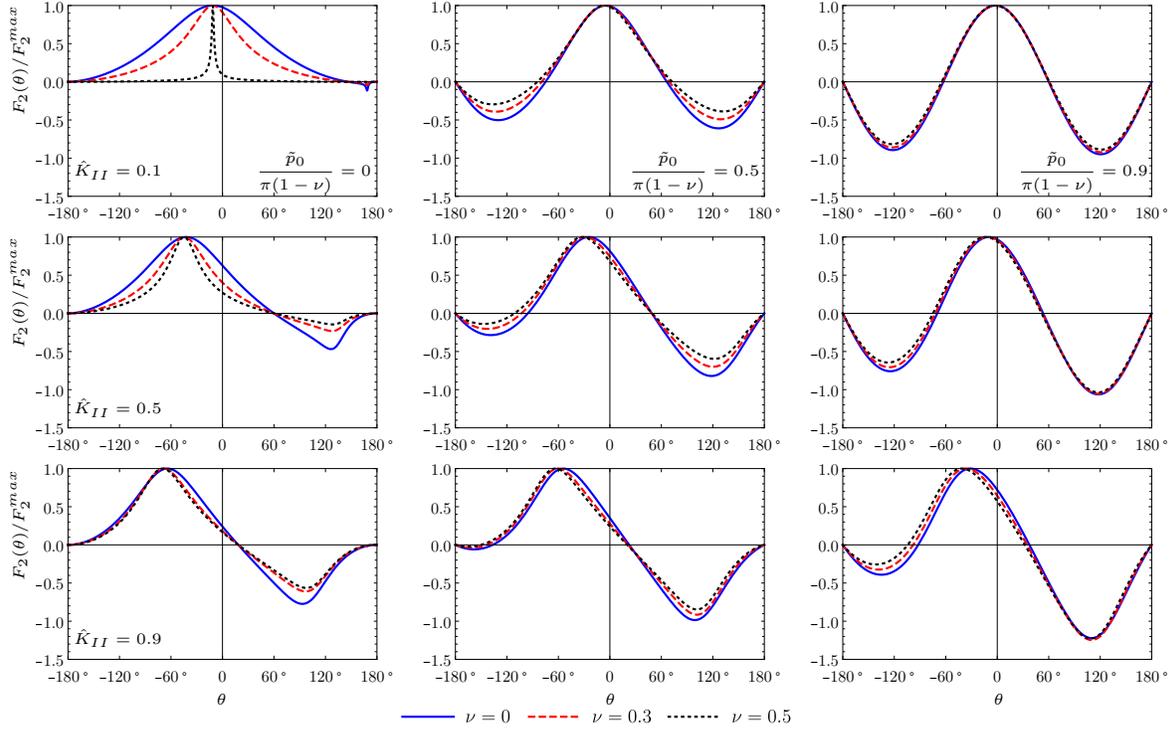

The graphs of function $F_2(\theta)$ normalised by its maximum value, $F_2^{\text{max}}$, are presented in Fig. \ref{MMCS_func} for three values of $\hat K_{II}=\{0.1,0.5,0.9\}$ and $\tilde p_0/(\pi(1-\nu))=\{0,0.5,0.9\}$. One can clearly see that this time, unlike the MDSED criterion, there is no problem with existence or uniqueness of the solution. There is only one maximum of $F_2$ in the interval $[-\pi, \pi]$. Moreover, this maximum corresponds by nature to a positive value of the circumferential stress, $\sigma_{\theta \theta}$. Relatively small influence of the Poisson's ratio on the crack propagation angle can be seen in the analyzed examples, however this issue will be discussed in more detail later on.

\begin{figure}[htb!]
\begin{center}
\includegraphics[scale=0.5]{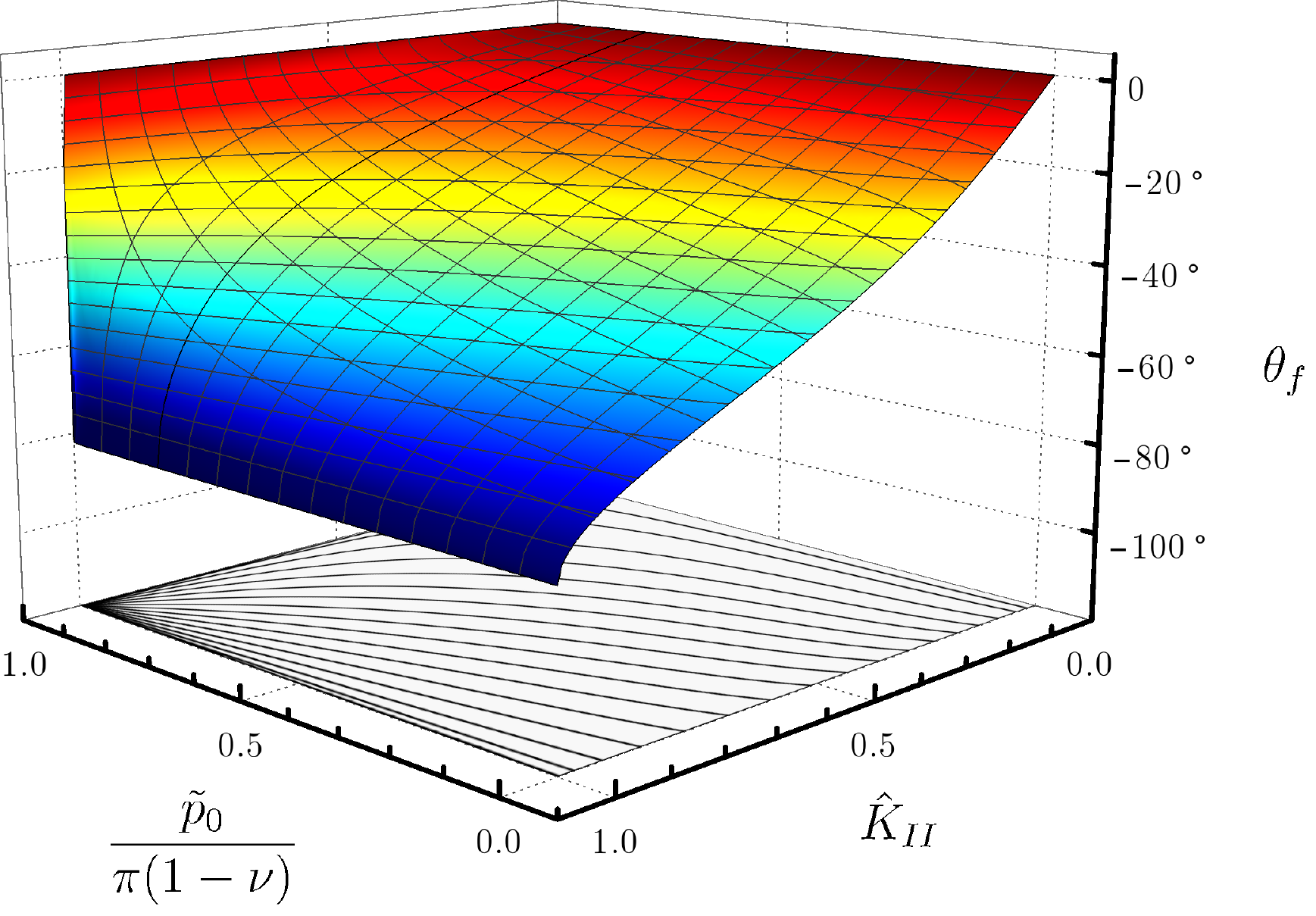}
\hspace{0mm}
\includegraphics[scale=0.45]{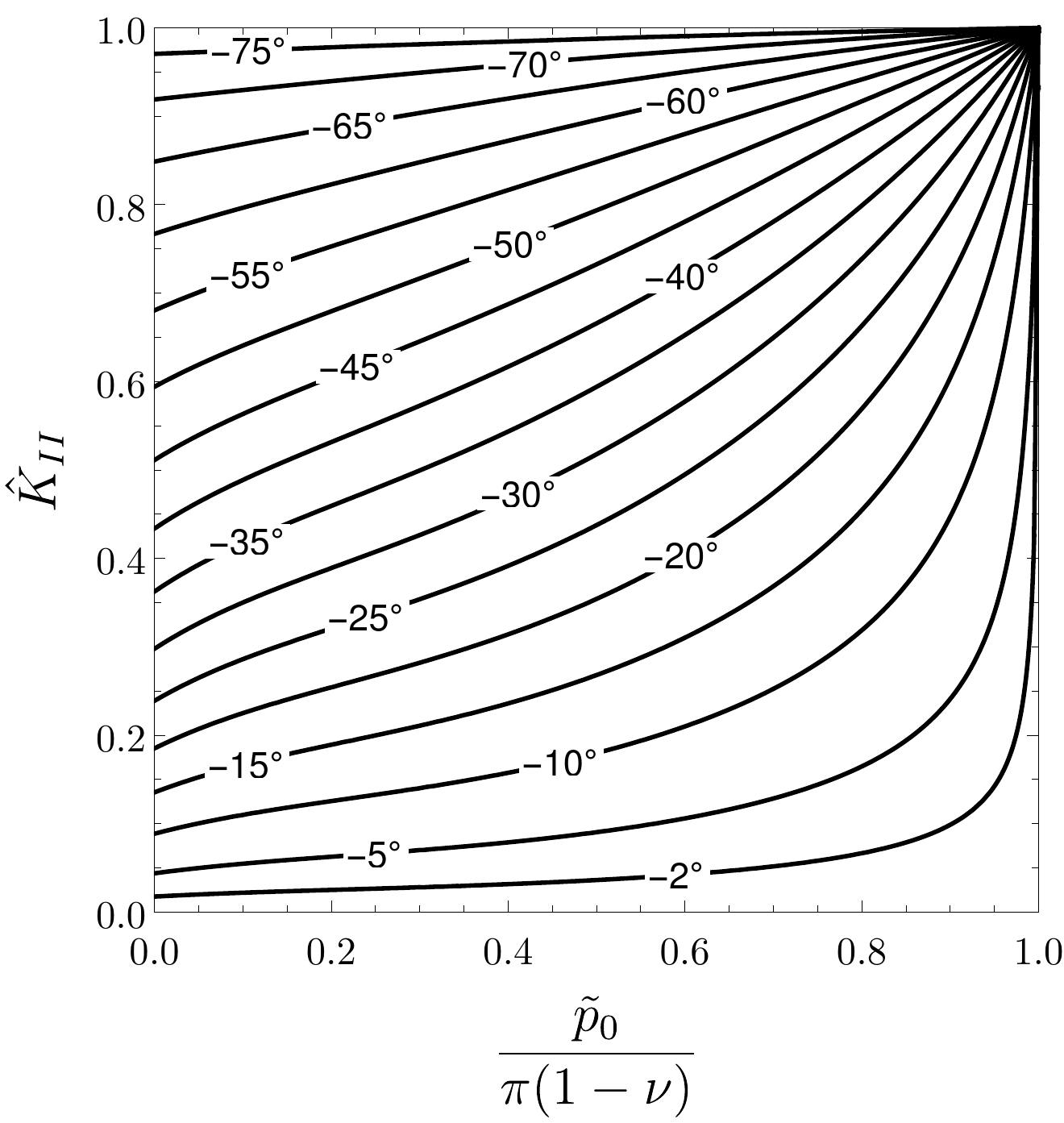}
\put(-165,20){$A$}
\put(5,20){$B$}
\put(-165,180){$D$}
\put(5,180){$C$}
\caption{MMCS-vM: Predicted propagation angle $\theta_f$ for $\hat K_{II} \in [0,1]$ and $\tilde p_0 \in [0,\pi(1-\nu)]$ for $\nu=0.3$. }
\label{MMCS_n03}
\end{center}
\end{figure}

In Fig. \ref{MMCS_n03} the solution, $\theta_f$, is presented for all admissible values of $\hat K_{II}$ and $\tilde p_0$ for a fixed Poisson's ratio, $\nu=0.3$. As can be seen, a continuous solution is obtained in almost the entire range of considered parameters (except for the point C). Naturally, for $\hat K_{II}=0$ one has $\theta_f=0$, however this time no fracture deflection is observed over the whole span of the viscosity dominated regime (edge BC). For $\hat K_{II}=1$ one has an invariable value of the crack propagation angle which amounts to -82.3$^\circ$. A peculiar behaviour of $\theta_f$ manifests in the corner C (viscosity dominated regime under severe antisymmetric load) where the solution has no limit. Likewise in the MDSED criterion, the actual angle of crack propagation depends here on the history of loading. As compared to the MDSED case, the MMCS-vM criterion provides lower values of $\theta_f$.

\begin{figure}[htb!]
\begin{center}
\includegraphics[scale=0.7]{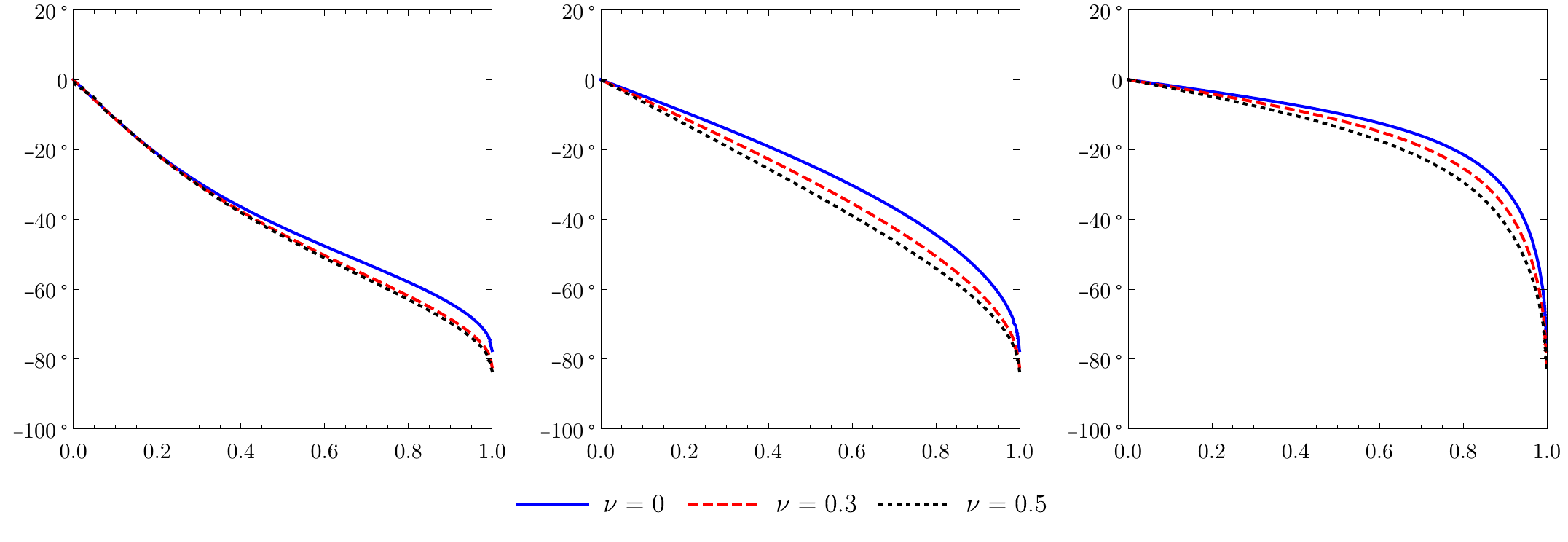}
{
\scriptsize
\put(-433,138){$\theta_f$}
\put(-287,138){$\theta_f$}
\put(-140,138){$\theta_f$}
\put(-360,17){$\hat K_{II}$}
\put(-214,17){$\hat K_{II}$}
\put(-66,17){$\hat K_{II}$}
\put(-350,133){$\ds \frac{\tilde p_0}{\pi(1-\nu)}=0$}
\put(-210,133){$\ds \frac{\tilde p_0}{\pi(1-\nu)}=0.5$}
\put(-64,133){$\ds \frac{\tilde p_0}{\pi(1-\nu)}=0.9$}
}
\caption{MMCS-vM: Value of $\theta_f$ for various values of Poisson's ratio.}
\label{MMCSvM_nu}
\end{center}
\end{figure}

Lastly, an impact of the Poisson's ratio on the propagation angle is given in Fig. \ref{MMCSvM_nu}. The discrepancies between results obtained for three analysed values of $\nu$ are comparable to the ones for the MCS criterion for $\tilde p_0>0$ (compare with Fig. 4b and Fig. 4c  in \cite{Perkowska_2017}) and can reach over $6^\circ$. However, in the toughness dominated regime ($\tilde p_0=0$, $\hat K_f=0$, Fig. \ref{MMCSvM_nu}a) a dependence of the solution on the Poisson's ratio can be noticed, which was not the case for the MCS approach (Fig. 4a in \cite{Perkowska_2017}). For $\hat K_{II}=0$ and $\hat K_{II}=1$ the solution does not depend on the Poisson's ratio.

\subsection{MMCS criterion: Drucker-Prager yield criterion (MMCS-DP)}
\label{sec:mmcsdp}

The second variant of the MMCS criterion utilizes the the Drucker-Prager yield theory in which the following relation holds:
\begin{equation}\label{dp}
\alpha I_1 + \sqrt{J_2} = \left( \alpha + \frac{1}{\sqrt{3}} \right) \sigma_t,
\end{equation}
where $I_1 = \tr \bsigma$, $J_2 = 1/2\, \dev \bsigma \cdot \dev \bsigma$, and $\alpha >0$ is a material parameter called pressure-sensitivity index and assumed to be a constant. It can be computed from experimental data by the following formula:
\begin{equation}\label{alphadp}
\alpha = \frac{\xi - 1}{\sqrt{3} (\xi + 1)}, \quad \xi = \frac{\sigma_c}{\sigma_t},
\end{equation}
where $\sigma_t$ and $\sigma_c$ are the yield stresses in uniaxial tension and compression, respectively.
Relation \eqref{alphadp} can be easily obtained by evaluating the yield function \eqref{dp} for a uniaxial compression state, $I_1 = -\sigma_c$, $J_2 = \sigma_c/\sqrt{3}$, which yields 
\begin{equation}
-\alpha \sigma_c + \frac{\sigma_c}{\sqrt{3}} = \left( \alpha + \frac{1}{\sqrt{3}} \right) \sigma_t, 
\end{equation}
and then solving for $\alpha$.

From \eqref{dp} one can derive an expression  for $r_b$:
\begin{equation}
\label{rb_vm}
r_b=R_b{\tilde r}_b(\theta), \quad \theta\in(-\pi,\pi),
\end{equation}
where
\begin{equation}
\label{R_b}
R_b=\frac{K_{IC}^2}{\sigma_t^2}, 
\end{equation}
has a dimension of length [m] and the dimensionless part is
\begin{align}\label{rb_DP}
\tilde{r}_b=&\frac{3 }{8 \pi (\sqrt{3} + 3\alpha)^2} \left[4\sqrt{3}
\alpha(1+\nu)\left(\hat K_I\cos\frac{\theta}{2}-\hat K_{II}\sin\frac{\theta}{2}\right)+ \right. \\
& \left. \sqrt{\hat K_I^2 c_1(\theta)+ \hat K_{II}^2 c_2(\theta)+ \hat K_I \hat K_{II} c_3(\theta)+
\hat K_f^2 c_4(\theta)+  \hat K_I \hat K_f c_5(\theta)+ \hat K_{II} \hat K_f c_6(\theta)+12\hat K_{III}^2}\right]^2,
\end{align}
where the functions $c_j(\theta)$ are defined in (\ref{constants_c}).
Note in the case of $\alpha=0$  the Drucker-Prager yield condition \eqref{dp} coincides with the von Mises condition \eqref{vm}. As for many geomaterials the tensile strength is 8-10 times smaller than their compressive strength \cite{Weinberger_1994}, we assume that  $\xi=10$ which yields $\alpha=0.4724$.
To compute the angle of crack propagation, one needs to find the maximum of an auxiliary function $F_2$ defined in \eqref{F2_def} (where $r_b$ is calculated from \eqref{rb_DP}) and apply condition \eqref{mmcsvm_sol}.

\begin{figure}[htb!]
\begin{center}
\includegraphics[scale=0.75]{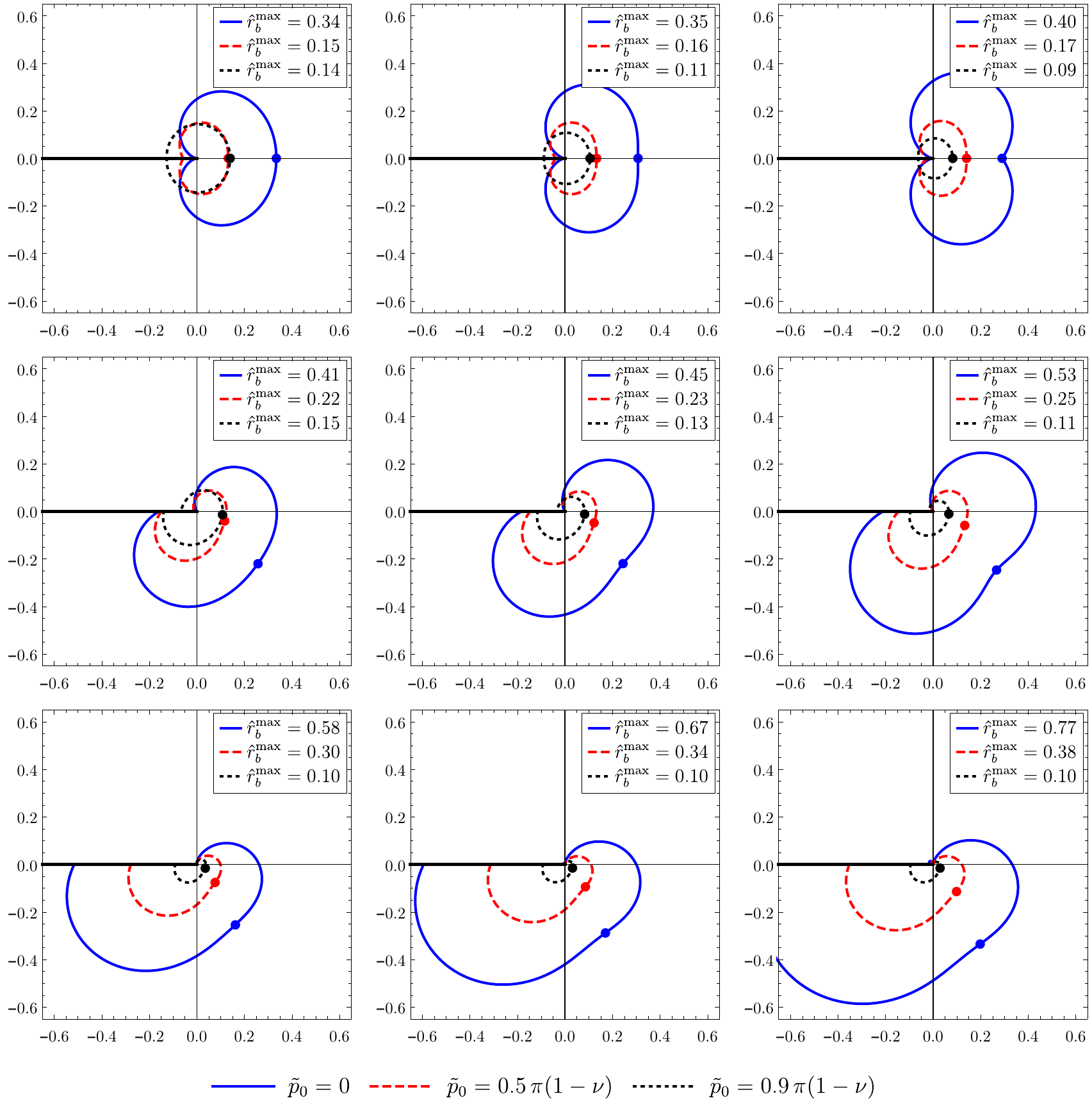}
{
\scriptsize
\put(-352,368){$\hat K_{II}=0$}
\put(-352,248){$\hat K_{II}=0.5$}
\put(-352,128){$\hat K_{II}=0.9$}
\put(-352,358){$\nu=0$}
\put(-224,368){$\nu=0.3$}
\put(-100,368){$\nu=0.5$}
}
\caption{MMCS-DP: The shapes of of the plastic zones described by the normalized radius $\hat r_b$ \eqref{r_hat} for various values of $\tilde p_0$ and fixed $\hat K_{II}$ and $\nu$. The blue lines reflect the actual sizes of the plastic zones, the red ones utilize the scaling factor 0.5, while the black curves are multiplied by 0.1.  The angle of crack propagation, $\theta_f$, is marked by a circle. }
\label{MMSCDP_rc}
\end{center}
\end{figure}
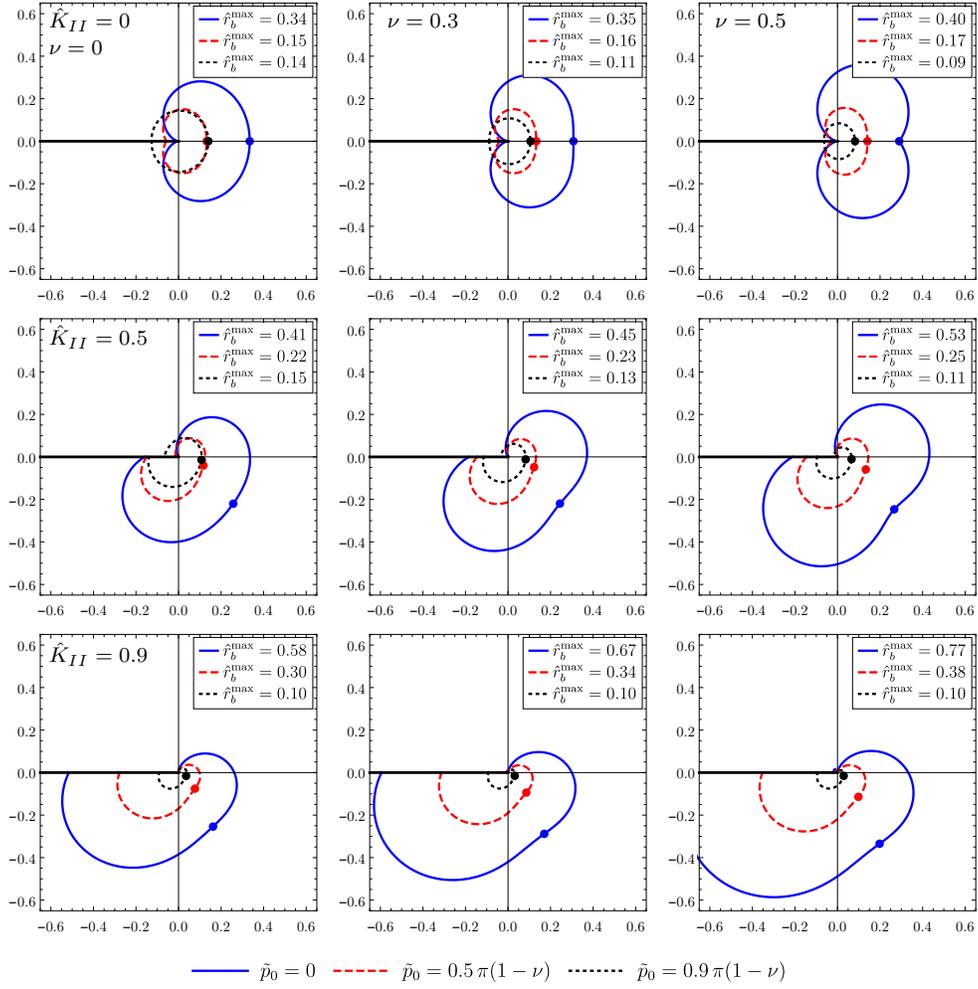

The graphs illustrating $ r_b(\theta)$ computed according to equation \eqref{rb_DP} are presented in Fig. \ref{MMSCDP_rc} (note that the scaling \eqref{r_hat} is used in the figure). This time we see that, contrary to the previous criteria based on the von Mises theory, the regular elliptic shape of the plastic zone when approaching the viscosity dominated regime is retained only for the symmetric loading. Indeed, it is the only case when any symmetry of the yield area (here with respect to the plane of crack propagation) is observed. When increasing the value of $\hat K_{II}$  the plastic zone shape becomes increasingly more irregular with a characteristic cut-off along the negative part of the horizontal axis.   As for the size of the yield area, it is always the greatest for the viscosity dominated regime and the smallest for the toughness dominated mode. Just as it was in the case of von Mises criterion, two different trends can be observed for the limiting regimes. In the viscosity dominated one the size of the plastic zone decreases with growing $\hat K_{II}$, while for the toughness dominated mode an inverse tendency holds. In the figure we can also see that for constant $\hat K_{II}$ an increase in $\tilde p_0$ reduces the angle of crack propagation.

\begin{figure}[htb!]
\begin{center}
\includegraphics[scale=0.7]{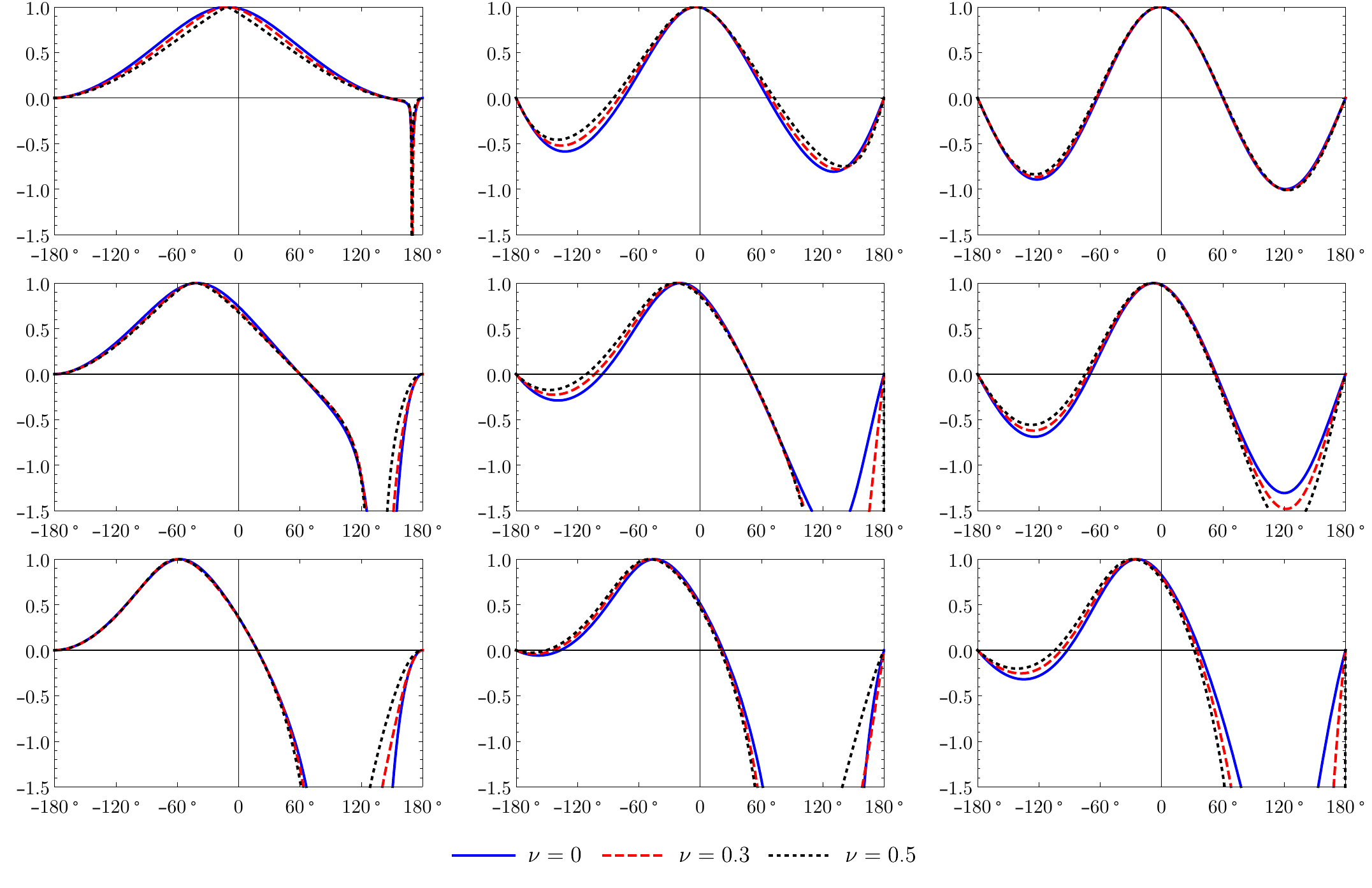}
{
\tiny
\put(-440,62){\rotatebox{90}{$\ds F_2(\theta)/F_2^{max}$}}
\put(-440,150){\rotatebox{90}{$\ds F_2(\theta)/F_2^{max}$}}
\put(-440,238){\rotatebox{90}{$\ds F_2(\theta)/F_2^{max}$}}
\put(-362,15){$\theta$}
\put(-215,15){$\theta$}
\put(-68,15){$\theta$}
\put(-416,215){$\hat K_{II}=0.1$}
\put(-416,125){$\hat K_{II}=0.5$}
\put(-416,38){$\hat K_{II}=0.9$}
\put(-347,215){$\ds \frac{\tilde p_0}{\pi(1-\nu)}=0$}
\put(-206,215){$\ds \frac{\tilde p_0}{\pi(1-\nu)}=0.5$}
\put(-59,215){$\ds \frac{\tilde p_0}{\pi(1-\nu)}=0.9$}
}
\caption{MMCS-DP: Value of $F_2(\theta)/F_2^{max}$ for various values of Poisson's ratio and fixed $\hat K_{II}$ and $\tilde p_0$. }
\label{MMCSDP_func}
\end{center}
\end{figure}

In Fig. \ref{MMCSDP_func} the values $F_2(\theta)/\max(F_2)$  over the interval $\theta \in [-\pi, \pi]$ are plotted. As before for the MMCS-VM criterion, no problems with existence or uniqueness of solution are reported. There always exists a single value of $\theta_f$ for which the maximum of $F_2$ (that corresponds to positive circumferential stress) is obtained. Again, one can observe a relatively low sensitivity of $\theta_f$ to the Poisson's ratio.

\begin{figure}[htb!]
\begin{center}
\includegraphics[scale=0.5]{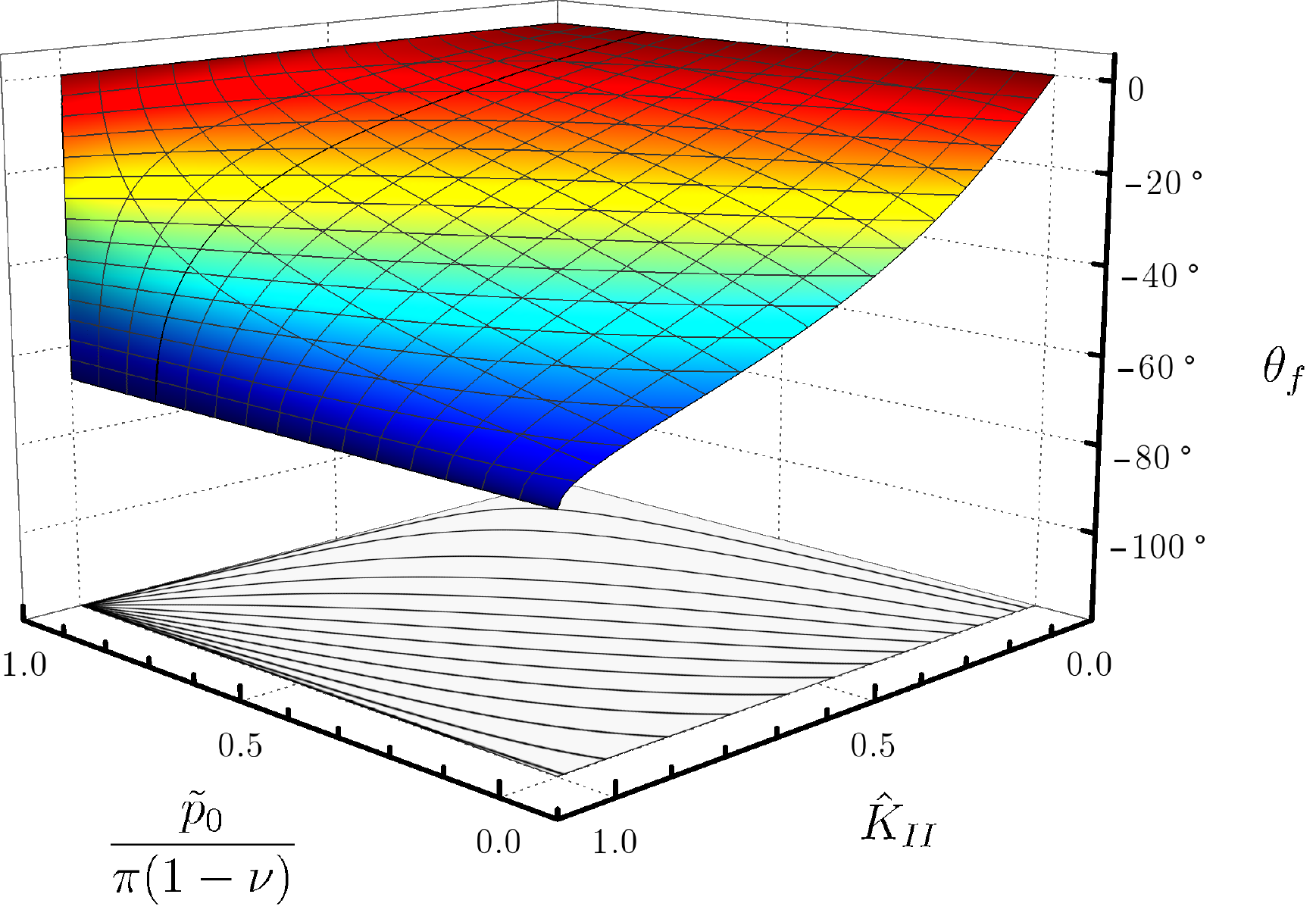}
\hspace{0mm}
\includegraphics[scale=0.45]{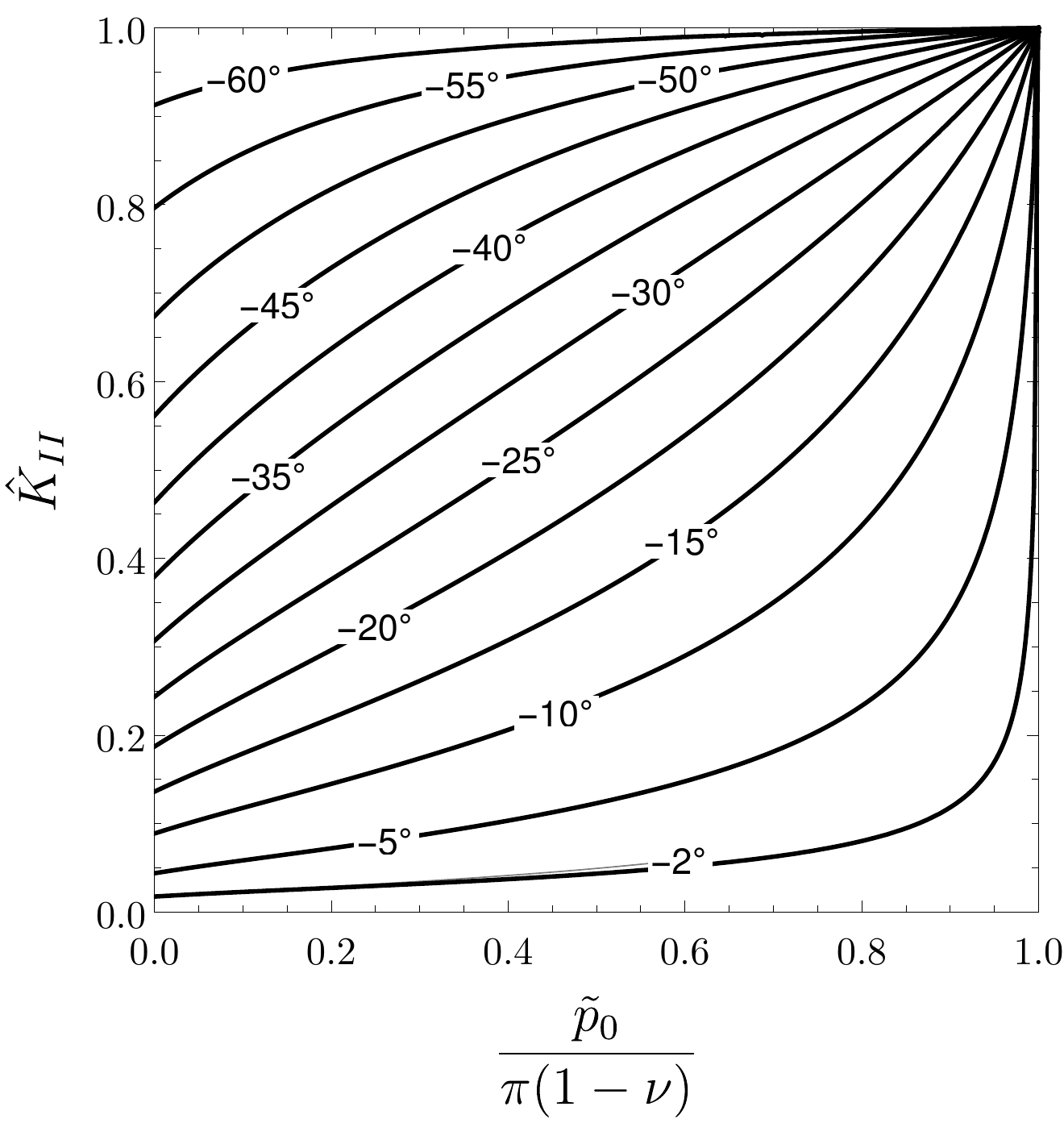}
\put(-165,20){$A$}
\put(5,20){$B$}
\put(-165,180){$D$}
\put(5,180){$C$}
\caption{MMCS-DP: Predicted propagation angle $\theta_f$ for $\hat K_{II} \in [0,1]$ and $\tilde p_0 \in [0,\pi(1-\nu)]$ for $\nu=0.3$. }
\label{MMSCDP_n03}
\end{center}
\end{figure}

In Fig. \ref{MMSCDP_n03} the solution, $\theta_f$, for all admissible values of $\hat K_{II} \in [0,1]$ and $\tilde p_0 \in [0,\pi(1-\nu)]$  is presented for $\nu=0.3$.  Similarly at it was in the MMCS variant utilizing von Mises yield criterion, a smooth solution exists over the whole range of the analyzed parameters. It is again only the point C  ($\tilde p_0=\pi (1-\nu)$, $\hat K_{II}=1$) where $\theta_f$ has no limit and its actual value depends on the loading history. Conventionally $\theta_f=0$ for $\hat K_{II}=0$. Moreover, no crack reorientation is observed in the viscosity dominated regime ($\tilde p_0=\pi (1-\nu))$. For a severe antisymmetric load ($\hat K_{II}=1$) the angle of crack propagation remains constant $\theta_f=-67.63^{\circ}$. In general, when using MMCS-DP criterion one obtains  smaller values of $\theta_f$ than those from MMCS-vM.

\begin{figure}[htb!]
\begin{center}
\includegraphics[scale=0.7]{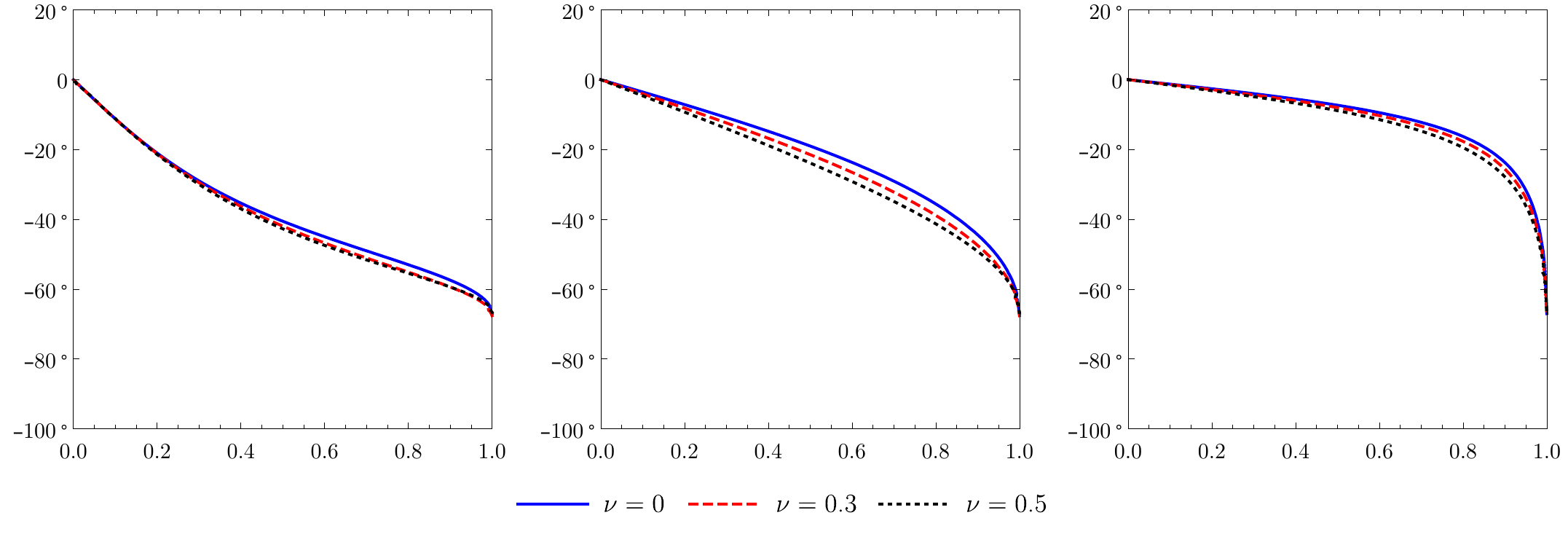}
{
\scriptsize
\put(-433,138){$\theta_f$}
\put(-287,138){$\theta_f$}
\put(-140,138){$\theta_f$}
\put(-360,17){$\hat K_{II}$}
\put(-214,17){$\hat K_{II}$}
\put(-66,17){$\hat K_{II}$}
\put(-350,133){$\ds \frac{\tilde p_0}{\pi(1-\nu)}=0$}
\put(-210,133){$\ds \frac{\tilde p_0}{\pi(1-\nu)}=0.5$}
\put(-64,133){$\ds \frac{\tilde p_0}{\pi(1-\nu)}=0.9$}
}
\caption{MMCS-DP: Value of $\theta_f$ for various values of Poisson's ratio and: a) $\frac{\tilde p_0}{\pi(1-\nu)}=0$, b) $\frac{\tilde p_0}{\pi(1-\nu)}=0.5$, c) $\frac{\tilde p_0}{\pi(1-\nu)}=0.9$.}
\label{MMCSDP_nu}
\end{center}
\end{figure}

Finally, to analyse how the value of Poisson's ratio affects the solution, graphs for three values of $\nu$ are plotted in Fig. \ref{MMCSDP_nu}. The solution for $\tilde p_0=0$ and $\tilde p_0=0.9\pi(1-\nu)$ depends on the value of $\nu$ to a very little extent. An impact of the Poisson's ratio was more noticeable for the von Mises yield criterion. Here,  as can be seen in the figure, the maximum deviation of results for the neighbouring values of $\nu$ does not exceed $4^{\circ}$.

\subsection{MMCS criterion: Tresca yield criterion (MMCS-TR)}
\label{sec:mmcstr}

Let us analyze now the problem of fracture deflection on the assumption that the Tresca criterion is used to define the yield stress. The criterion itself reads \cite{Bigoni_2004}:
\begin{equation}\label{tr}
2 \sqrt{J_2} \cos \left[ \frac{1}{3} \cos^{-1} \left( \frac{3\sqrt{3} J_3}{2 J_2^{3/2}} \right) - \frac{\pi}{6} \right] = \sigma_t,
\end{equation}
where $J_3 = 1/3\, \tr (\dev \bsigma)^3$ and $\sigma_t$ is the uniaxial yield stress. Conventionally, the angle of crack propagation is found by maximizing the function $F_2$ \eqref{F2_def} along the elastic-plastic boundary. This time we do not write an expression for $r_b$ as it is to complex.

\begin{figure}[htb!]
\begin{center}
\includegraphics[scale=0.75]{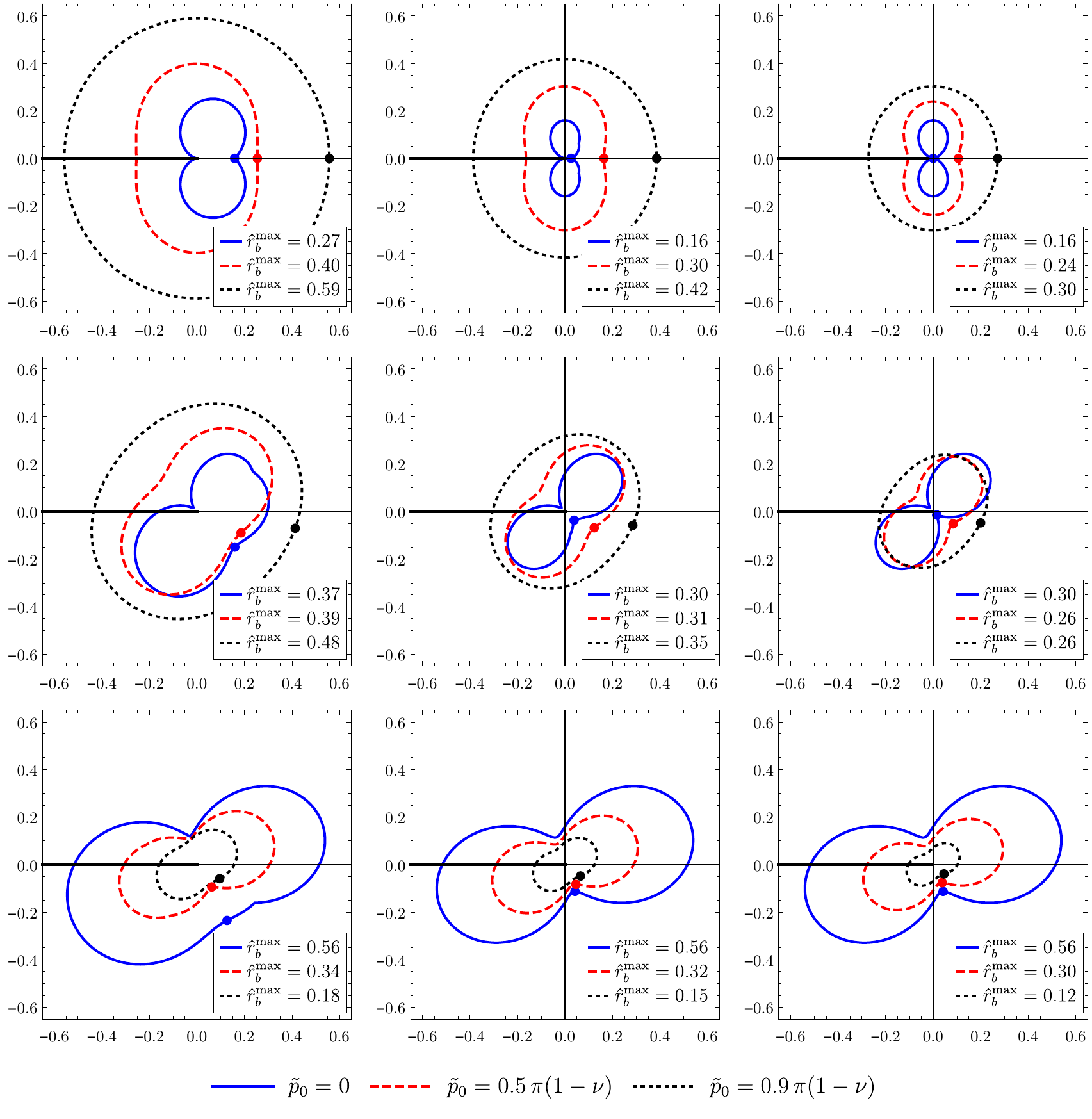}
{
\scriptsize
\put(-352,368){$\hat K_{II}=0$}
\put(-352,248){$\hat K_{II}=0.5$}
\put(-352,128){$\hat K_{II}=0.9$}
\put(-274,368){$\nu=0$}
\put(-156,368){$\nu=0.3$}
\put(-33,368){$\nu=0.5$}
}
\caption{MMCS-TR: The shapes of of the plastic zones described by the normalized radius $\hat r_b$ \eqref{r_hat} for various values of $\tilde p_0$ and fixed $\hat K_{II}$ and $\nu$. The blue lines reflect the actual sizes of the plastic zones, the red ones utilize the scaling factor 0.5, while the black curves are multiplied by 0.1.  The angle of crack propagation, $\theta_f$, is marked by a circle.}
\label{MMSCTR_rc}
\end{center}
\end{figure}

The resulting areas of the plastic yield are depicted in Fig. \ref{MMSCTR_rc}. As can be seen, the obtained shapes are very similar to those for the MDSED criterion (compare Fig. \ref{MDSED_rc}). The general trends remain the same as in the recalled case, however the zones of plastic yield are larger now (which is quite obvious if one notes that the von Mises yield stress is always greater or equal to the Tresca yield stress). Some irregularities of the boundary are observed in the viscosity dominated regime for $\hat K_{II}=0.5$ and $\hat K_{II}=0.9$, which shall be explained later on.

\begin{figure}[htb!]
\begin{center}
\includegraphics[scale=0.7]{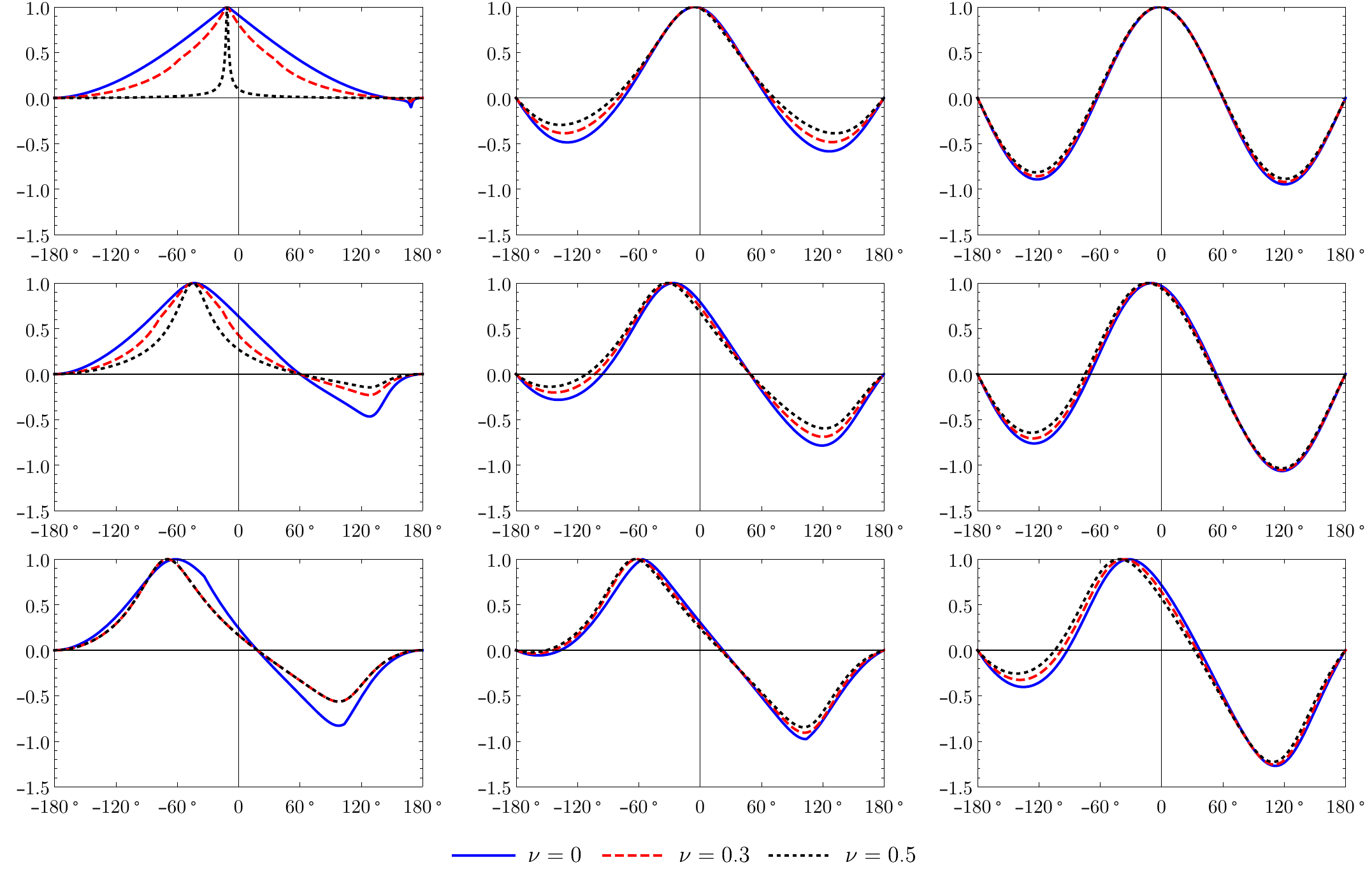}
{
\tiny
\put(-440,62){\rotatebox{90}{$\ds F_2(\theta)/F_2^{max}$}}
\put(-440,150){\rotatebox{90}{$\ds F_2(\theta)/F_2^{max}$}}
\put(-440,238){\rotatebox{90}{$\ds F_2(\theta)/F_2^{max}$}}
\put(-362,15){$\theta$}
\put(-215,15){$\theta$}
\put(-68,15){$\theta$}
\put(-416,215){$\hat K_{II}=0.1$}
\put(-416,125){$\hat K_{II}=0.5$}
\put(-416,38){$\hat K_{II}=0.9$}
\put(-347,215){$\ds \frac{\tilde p_0}{\pi(1-\nu)}=0$}
\put(-206,215){$\ds \frac{\tilde p_0}{\pi(1-\nu)}=0.5$}
\put(-59,215){$\ds \frac{\tilde p_0}{\pi(1-\nu)}=0.9$}
}
\caption{MMCS-TR: Value of $F_2(\theta)/F_2^{max}$ for various values of Poisson's ratio and fixed $\hat K_{II}$ and $\tilde p_0$.}
\label{MMCSTR_func}
\end{center}
\end{figure}

In Fig. \ref{MMCSTR_func} the qualitative behaviour of function $F_2(\theta)$ is depicted.  It resembles to a large degree that obtained for the MMCS-vM criterion (compare Fi.g \ref{MMCS_func}). Again, no problem with existence or uniqueness of solution appears.

\begin{figure}[htb!]
\begin{center}
\includegraphics[scale=0.5]{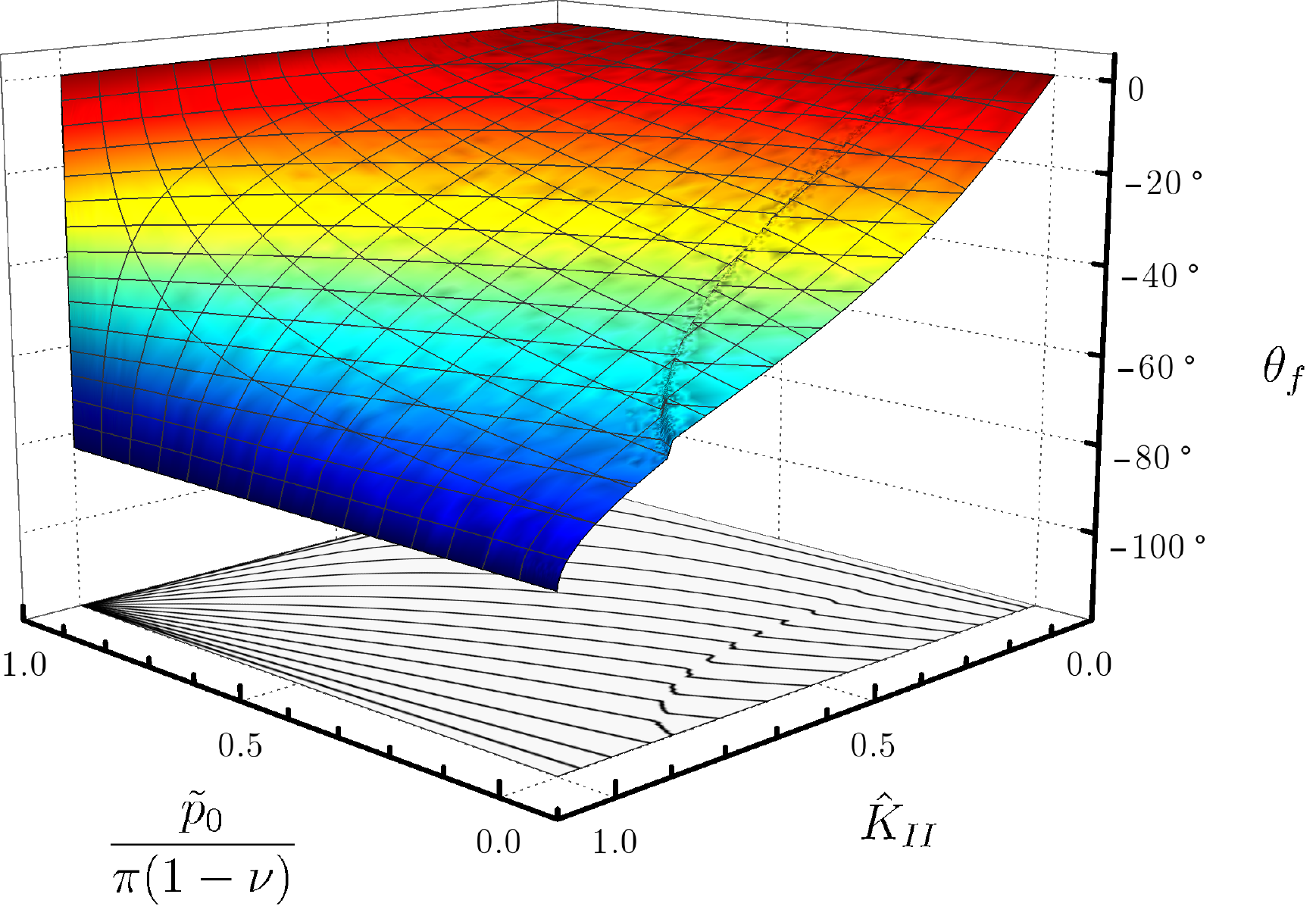}
\hspace{0mm}
\includegraphics[scale=0.45]{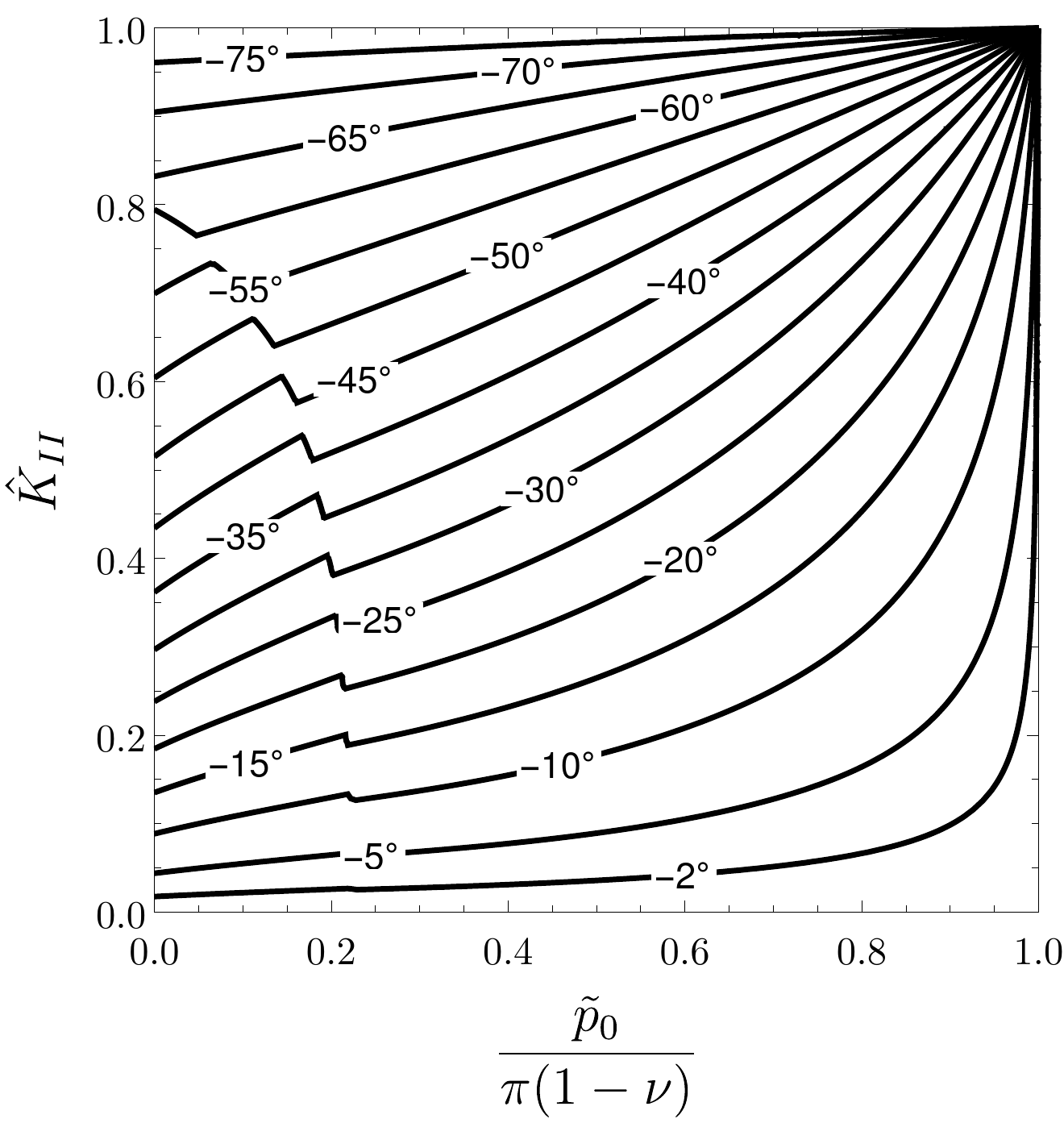}
\put(-165,20){$A$}
\put(5,20){$B$}
\put(-165,180){$D$}
\put(5,180){$C$}
\caption{MMCS-TR: Predicted propagation angle $\theta_f$ for $\hat K_{II} \in [0,1]$ and $\tilde p_0 \in [0,\pi(1-\nu)]$ for $\nu=0.3$.}
\label{MMSCTR_n03}
\end{center}
\end{figure}

\begin{figure}[htb!]
\begin{center}
\includegraphics[scale=0.7]{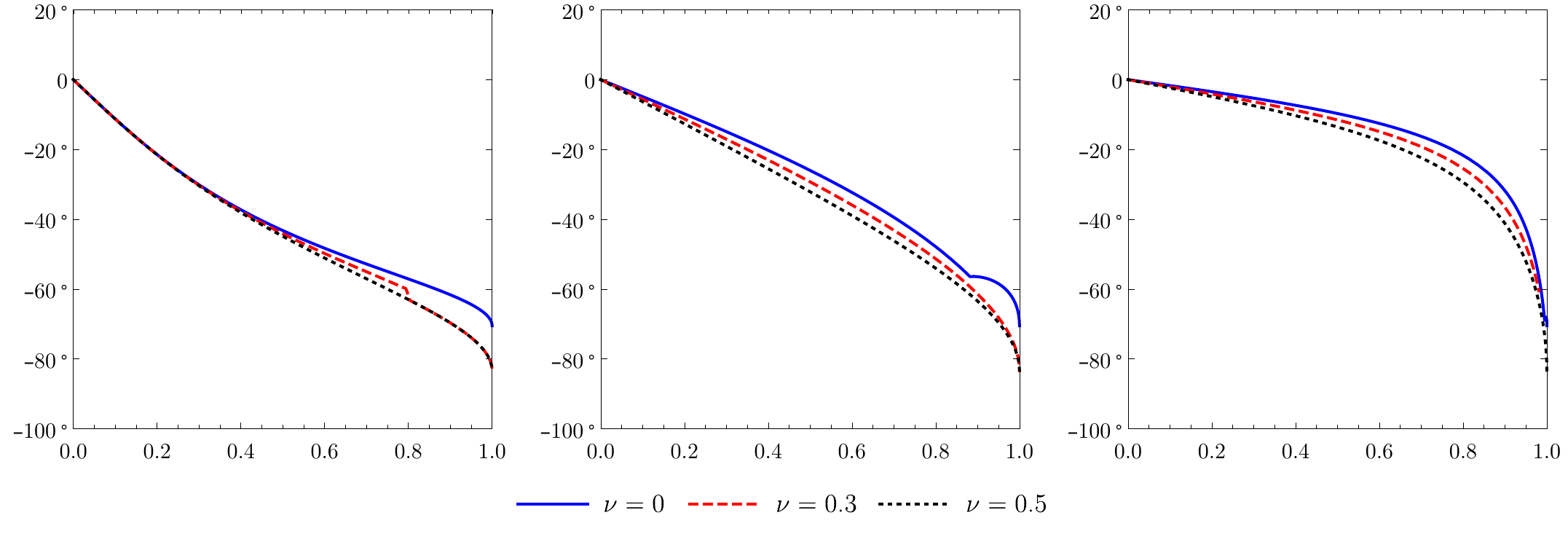}
{
\scriptsize
\put(-433,138){$\theta_f$}
\put(-287,138){$\theta_f$}
\put(-140,138){$\theta_f$}
\put(-360,17){$\hat K_{II}$}
\put(-214,17){$\hat K_{II}$}
\put(-66,17){$\hat K_{II}$}
\put(-350,133){$\ds \frac{\tilde p_0}{\pi(1-\nu)}=0$}
\put(-210,133){$\ds \frac{\tilde p_0}{\pi(1-\nu)}=0.5$}
\put(-64,133){$\ds \frac{\tilde p_0}{\pi(1-\nu)}=0.9$}
}
\caption{MMCS-TR: Value of $\theta_f$ for various values of Poisson's ratio and: a) $\frac{\tilde p_0}{\pi(1-\nu)}=0$, b) $\frac{\tilde p_0}{\pi(1-\nu)}=0.5$, c) $\frac{\tilde p_0}{\pi(1-\nu)}=0.9$.}
\label{MMCSTR_nu}
\end{center}
\end{figure}

Fig. \ref{MMSCTR_n03} presents a distribution of $\theta_f$ over the permissible range of $\tilde p_0$ and $\hat K_{II}$ for $\nu=0.3$. The general behaviour and values of solution are very similar to those reported for the MMCS-vM criterion (compare Fig. \ref{MMCS_n03}). Again no limit of $\theta_f$ is observed for $\tilde p_0 \to \pi (1-\nu)$ and $\hat K_{II}=1$ (corner C). For the extreme antisymmetric load ($\hat K_{II}=1$)  $\theta_f=-83.4^\circ$ was obtained. This time however we see some pecularity of the crack propagation angle. Namely there exists a line in the analyzed parametric space along which the solution smoothness is broken (it can be easily identified in the Fig. \ref{MMSCTR_n03}b)). After a careful analysis it turned out that this line corresponds to a situation when for the predefined mixed mode loading the edge of the Tresca yield surface is reached. Respective transition points can be also identified in Fig. \ref{MMCSTR_nu}a) and Fig. \ref{MMCSTR_nu}b). Moreover, the aforementioned mechanism manifests itself in the irregularities of the plastic zone shapes that were noted when discussing Fig. \ref{MMSCTR_rc}.

Finally, in Fig. \ref{MMCSTR_nu} the influence of the Poisson's ratio on the crack propagation angle is presented. It shows that the level of deviations between the results for the neighbouring values of $\nu$ is similar to that in the previous criteria. However, unlike the previous cases, one obtains coincidence of the respective results for $\hat K_{II}=1$ only when approaching the viscosity dominated regime. It is the variant of $\nu=0$ which increasingly diverges from the two remaining curves for declining $\tilde p_0$.

\subsection{MMCS criterion: Mohr-Coulomb yield criterion (MMCS-MC)}
\label{sec:mmcscm}

The last analyzed case assumes that the yield stress follows from the Mohr-Coulomb criterion \cite{Bigoni_2004}:

\begin{equation}\label{cm}
\frac{\xi - 1}{\sqrt{\xi^2 + \xi + 1}} \frac{I_1}{\sqrt{3}} +
2 \sqrt{J_2} \cos \left[ \frac{1}{3} \cos^{-1} \left( \frac{3\sqrt{3} J_3}{2 J_2^{3/2}} \right) - \tan^{-1} \left( \frac{\sqrt{3}}{2\xi + 1} \right) \right] = \frac{\sqrt{3} \sigma_c}{\sqrt{\xi^2 + \xi + 1}},
\end{equation}
where $\sigma_t$ and $\sigma_c$ are the yield stresses in uniaxial tension and compression, respectively, and $\xi = \sigma_c/\sigma_t$.
Throughout this subsection we use $\xi=10$, which is consistent with the value  $\alpha = 0.4724$ in the Drucker-Prager criterion in Sec.~\ref{sec:mmcsdp}. The angle of crack propagation is found as previously by maximizing the function $F_2$ \eqref{F2_def}. Again, as respective expression for the radius of the plastic zone is very complex, we do no present it here.

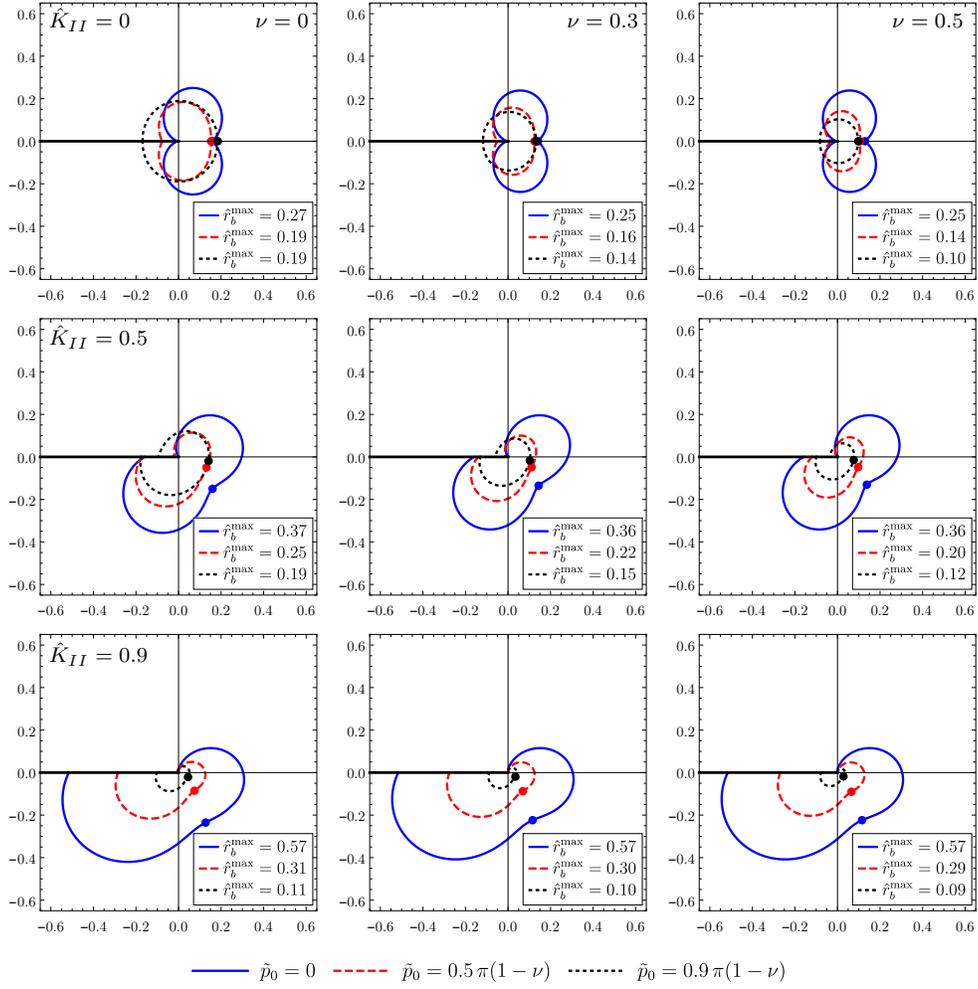
\begin{figure}[htb!]
\begin{center}
\includegraphics[scale=0.75]{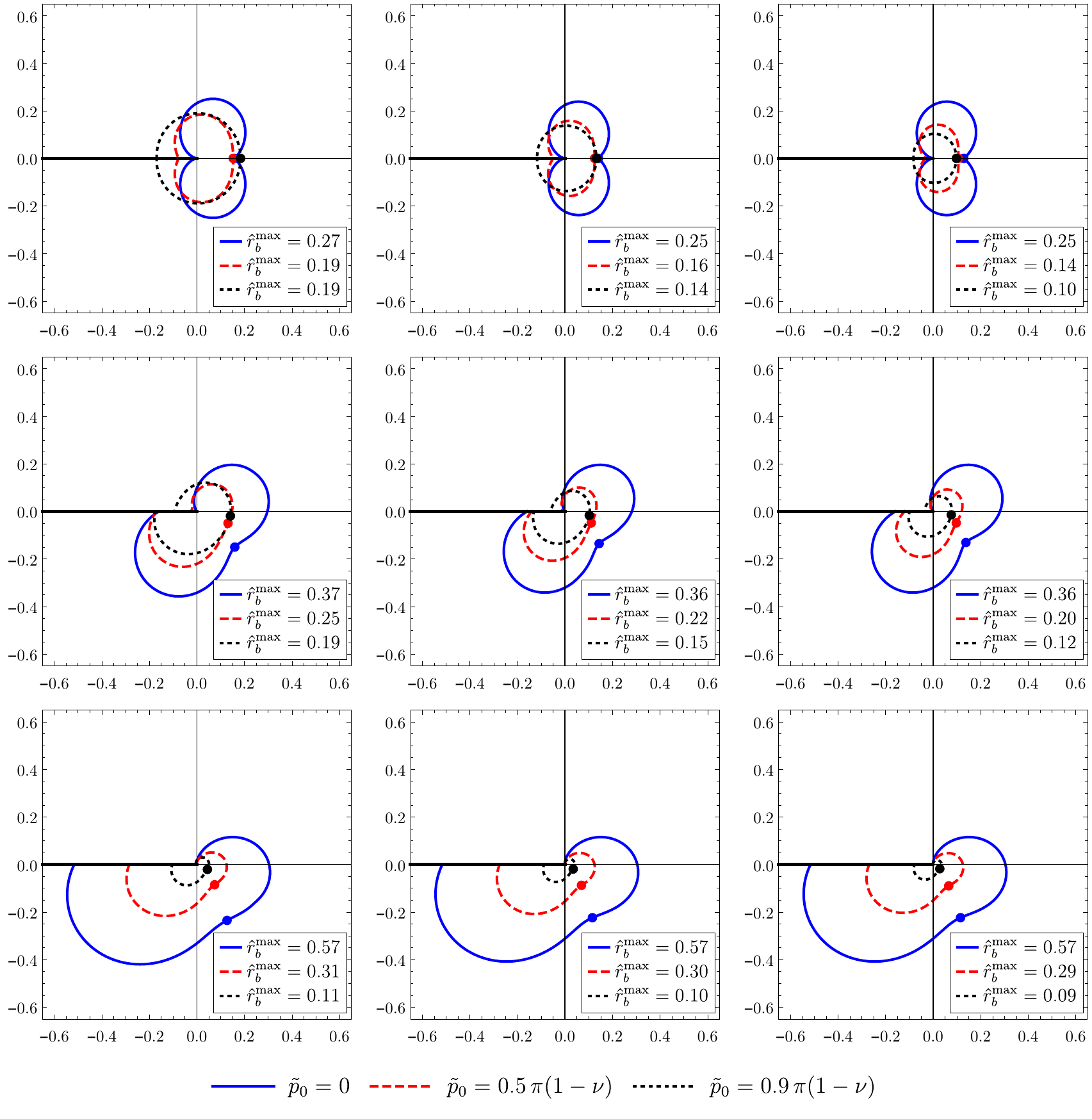}
{
\scriptsize
\put(-352,368){$\hat K_{II}=0$}
\put(-352,248){$\hat K_{II}=0.5$}
\put(-352,128){$\hat K_{II}=0.9$}
\put(-274,368){$\nu=0$}
\put(-156,368){$\nu=0.3$}
\put(-33,368){$\nu=0.5$}
}
\caption{MMCS-MC: The shapes of of the plastic zones described by the normalized radius $\hat r_b$ \eqref{r_hat} for various values of $\tilde p_0$ and fixed $\hat K_{II}$ and $\nu$. The blue lines reflect the actual sizes of the plastic zones, the red ones utilize the scaling factor 0.5, while the black curves are multiplied by 0.1.  The angle of crack propagation, $\theta_f$, is marked by a circle. }
\label{MMSCCM_rc}
\end{center}
\end{figure}

In Fig. \ref{MMSCCM_rc} the shapes of the plastic zones for different values of the analyzed parameters are shown. One can see a  resemblance of the contours of the zones to those obtained for the MMCS-DP criterion (compare Fig. \ref{MMSCDP_rc}). Their sizes are always greater for the  Drucker-Prager criterion, which is especially pronounced  in the toughness dominated regime ($\tilde p_0=0$). The general trends remain the same as in the MMCS-DP variant.

\begin{figure}[htb!]
\begin{center}
\includegraphics[scale=0.7]{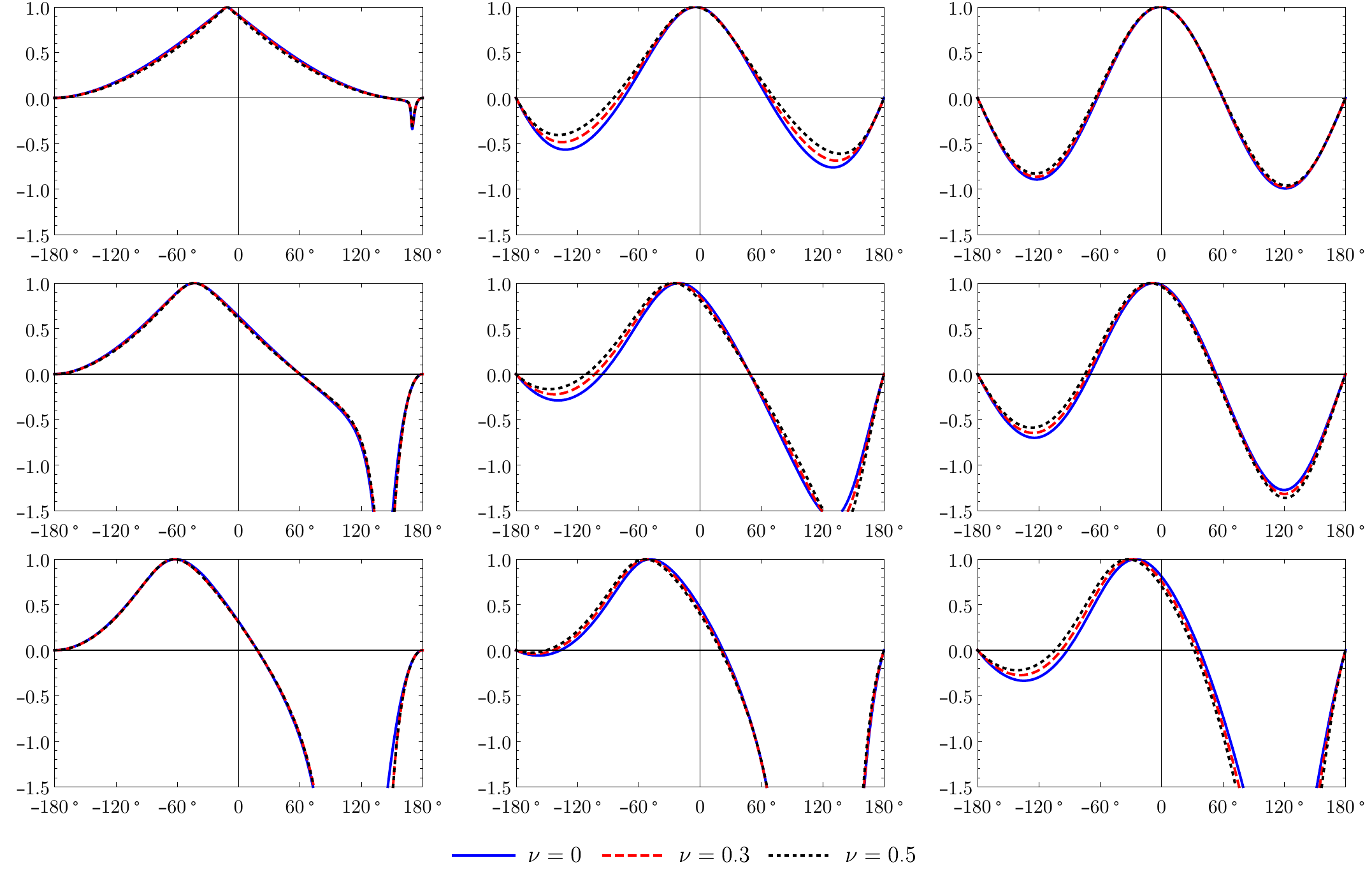}
{
\tiny
\put(-440,62){\rotatebox{90}{$\ds F_2(\theta)/F_2^{max}$}}
\put(-440,150){\rotatebox{90}{$\ds F_2(\theta)/F_2^{max}$}}
\put(-440,238){\rotatebox{90}{$\ds F_2(\theta)/F_2^{max}$}}
\put(-362,15){$\theta$}
\put(-215,15){$\theta$}
\put(-68,15){$\theta$}
\put(-416,215){$\hat K_{II}=0.1$}
\put(-416,125){$\hat K_{II}=0.5$}
\put(-416,38){$\hat K_{II}=0.9$}
\put(-347,215){$\ds \frac{\tilde p_0}{\pi(1-\nu)}=0$}
\put(-206,215){$\ds \frac{\tilde p_0}{\pi(1-\nu)}=0.5$}
\put(-59,215){$\ds \frac{\tilde p_0}{\pi(1-\nu)}=0.9$}
}
\caption{MMCS-MC: Value of $F_2(\theta)/F_2^{max}$ for various values of Poisson's ratio and fixed $\hat K_{II}$ and $\tilde p_0$. }
\label{MMCSCM_func}
\end{center}
\end{figure}
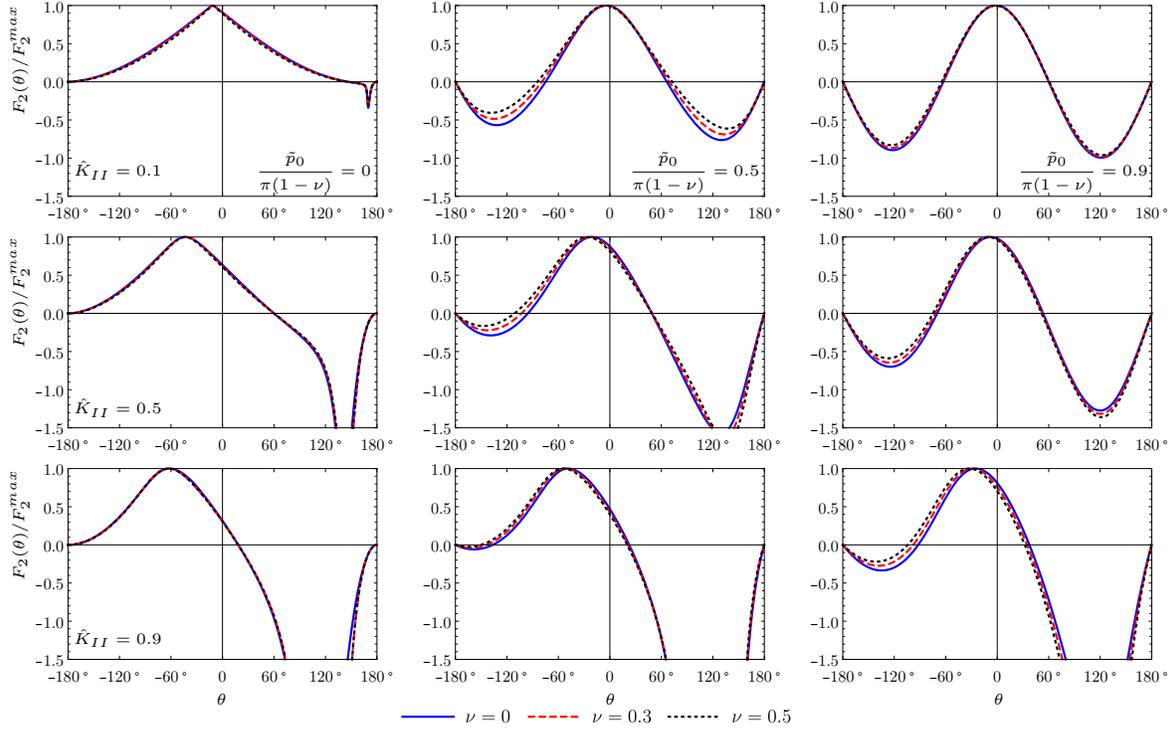

Fig. \ref{MMCSCM_func} shows the qualitative distribution of the $F_2$ function. We can see that the results are hardly distinguishable from those obtained for the  MMCS-DP. For every analyzed combination of parameters there exists a unique solution to the problem.

\begin{figure}[htb!]
\begin{center}
\includegraphics[scale=0.5]{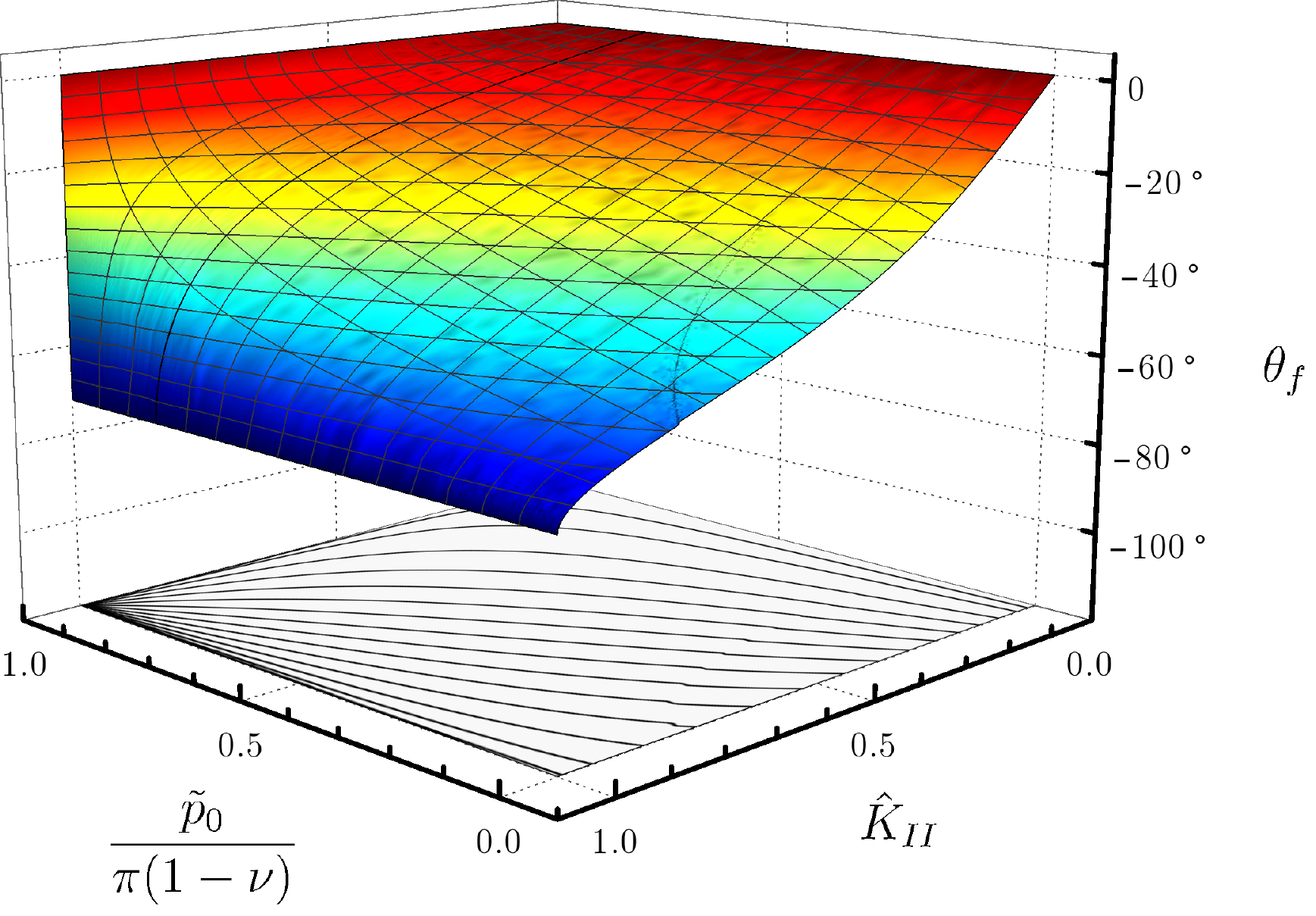}
\hspace{0mm}
\includegraphics[scale=0.45]{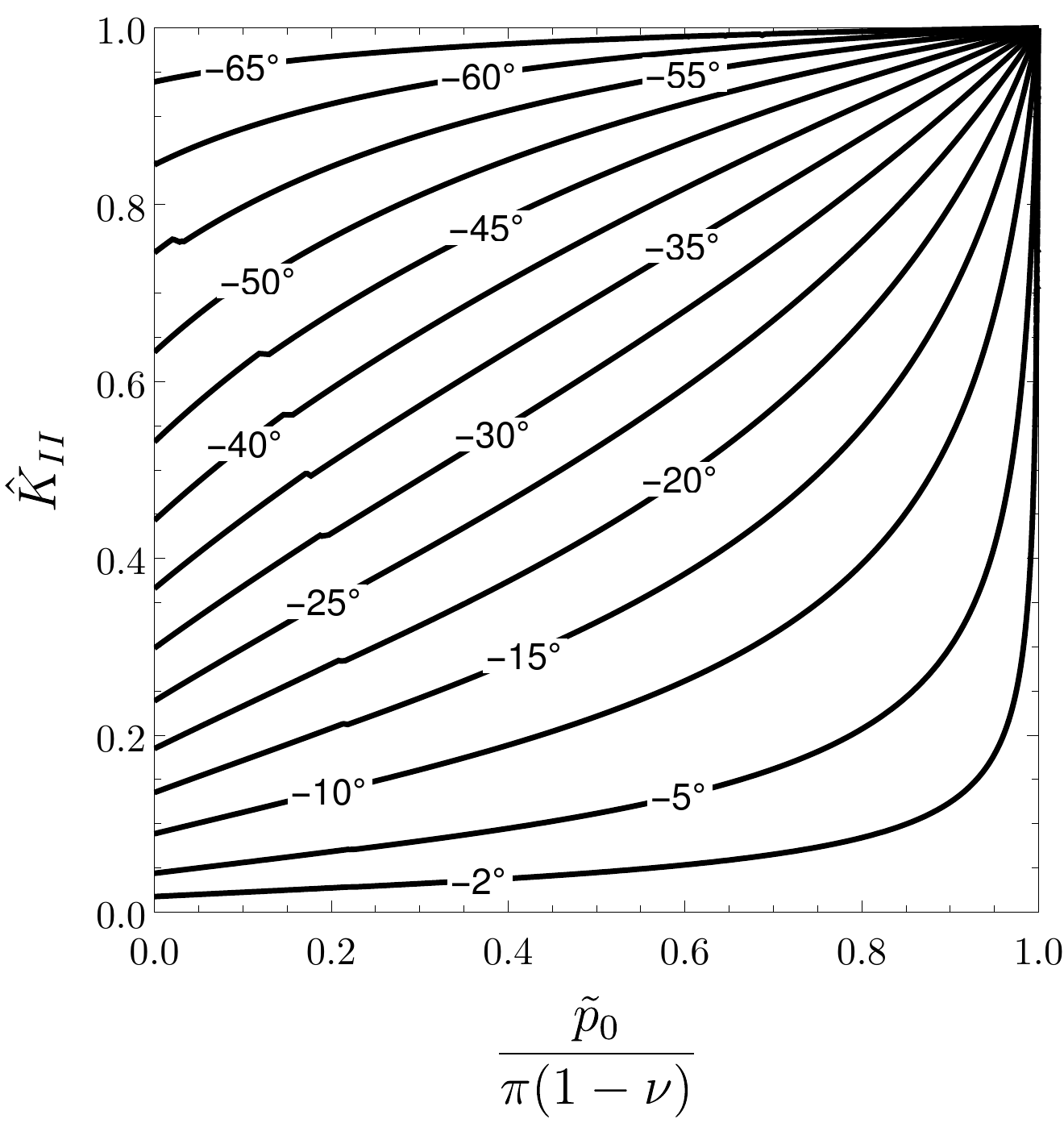}
\put(-165,20){$A$}
\put(5,20){$B$}
\put(-165,180){$D$}
\put(5,180){$C$}
\caption{MMCS-MC: Predicted propagation angle $\theta_f$ for $\hat K_{II} \in [0,1]$ and $\tilde p_0 \in [0,\pi(1-\nu)]$ for $\nu=0.3$.}
\label{MMSCCM_n03}
\end{center}
\end{figure}

A distribution of the crack propagation angle, $\theta_f$, over  $\hat K_{II} \in [0,1]$ and $\tilde p_0 \in [0,\pi(1-\nu)]$ for $\nu=0.3$ is shown in Fig. \ref{MMSCCM_n03}. The values of $\theta_f$ are  close to those computed for the MMCS-DP criterion. The general tendencies are also very similar. For $K_{II}=1$ $\theta_f=-72.45^\circ$ is obtained. Again there is a line in the analyzed parametric space (Fig. \ref{MMSCCM_n03}a)) along which the smoothness  of fracture deflection angle is broken. It is due to the edge of the Mohr-Coulomb yield surface, however the trend itself is much less pronounced than it was in the MMCS-DP.

\begin{figure}[htb!]
\begin{center}
\includegraphics[scale=0.7]{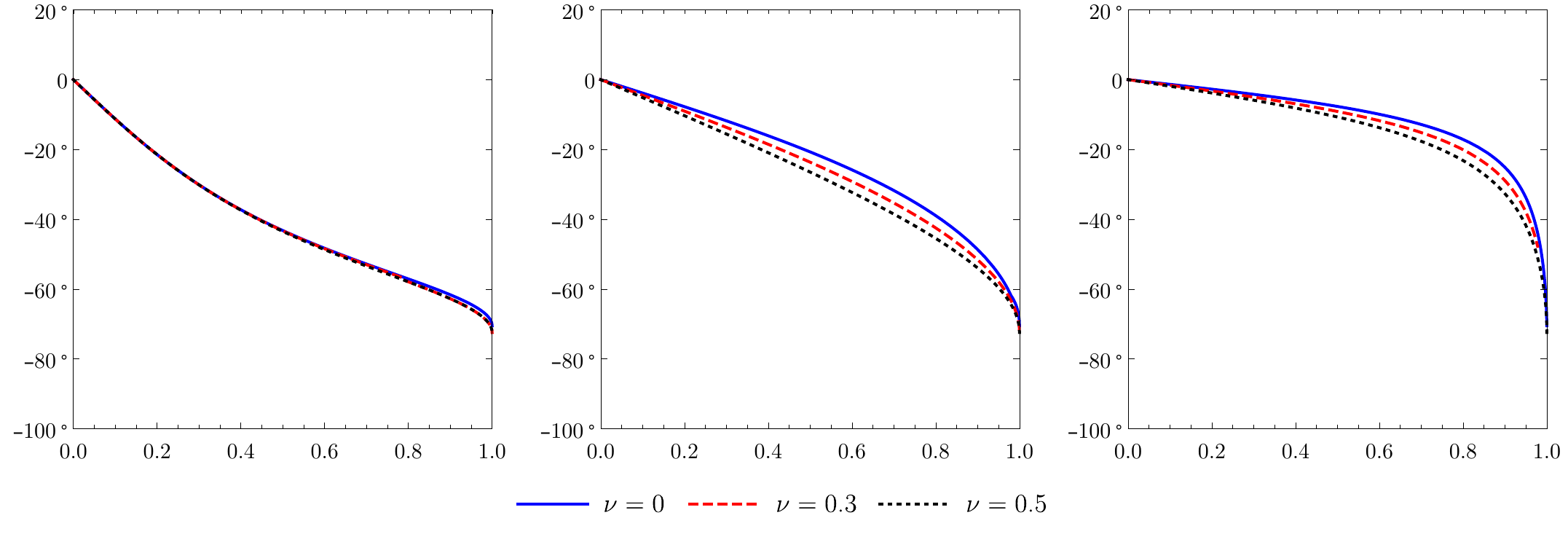}
{
\scriptsize
\put(-433,138){$\theta_f$}
\put(-287,138){$\theta_f$}
\put(-140,138){$\theta_f$}
\put(-360,17){$\hat K_{II}$}
\put(-214,17){$\hat K_{II}$}
\put(-66,17){$\hat K_{II}$}
\put(-350,133){$\ds \frac{\tilde p_0}{\pi(1-\nu)}=0$}
\put(-210,133){$\ds \frac{\tilde p_0}{\pi(1-\nu)}=0.5$}
\put(-64,133){$\ds \frac{\tilde p_0}{\pi(1-\nu)}=0.9$}
}
\caption{MMCS-MC: Value of $\theta_f$ for various values of Poisson's ratio and: a) $\frac{\tilde p_0}{\pi(1-\nu)}=0$, b) $\frac{\tilde p_0}{\pi(1-\nu)}=0.5$, c) $\frac{\tilde p_0}{\pi(1-\nu)}=0.9$.}
\label{MMCSCM_nu}
\end{center}
\end{figure}
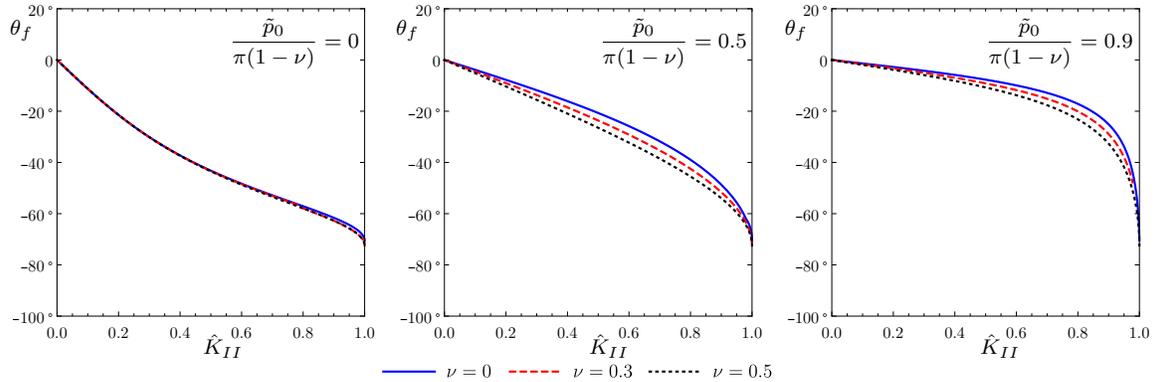

Finally, in Fig. \ref{MMCSCM_nu} the impact of the Poisson's ratio on $\theta_f$ is analyzed. Respective trends remain the same as in the MMCS-DP criterion. In the toughness dominated regime ($\tilde p_0=0$) the influence of $\nu$ is infinitesimal, respective curves are hardly distinguishable.

\section{Comparison between criteria}
\label{sec:comparison}

{In this section we shall provide a brief comparison of the results obtained for respective criteria. We decided to omit the MDSED criterion. This is because it either provides no solution in a relatively wide area near the viscosity dominated regime (MDSED - variant I) or the solution in this region exhibits some instability (MDSED - variant II). For this reason we deem the MDSED criterion questionable, at least in the recalled zone.

As for the remaining criteria we notice very strong similarities between the angles of crack propagation computed for the following pairs: i) MMCS-vM and MMCS-TR, ii) MMCS-DP and MMCS-MC. Its is not a surprise if one recalls that the von Mises yield surface can be considered a smooth approximation of the one for the Tresca criterion. Similarly, the Drucker-Prager yield surface is a smooth extension of that defined by the Mohr-Coulomb theory. Clearly, in those cases when the yield surface is represented by the quadratic in the space of principal stresses (MMCS-vM and MMCS-DP) the solution ($\theta_f$) is always a smooth function. Otherwise (MMCS-TR and MMCS-MC), if for a certain configuration of the loading the edge of the yield surface is transgressed, the angle of crack propagation is no longer smooth. This constitutes an evident drawback of the latter criteria.

In Fig. \ref{FigCOMP} - Fig. \ref{Fig2COMP} we have collated the results (the angles of crack propagation) obtained for the discussed criteria under different values of $\hat K_{II}$, $\tilde p_o$ and $\nu$. It shows that there is a good convergence between the respective curves representing $\theta_f$ for two groups of criteria: i)  MMCS-vM and MMCS-TR, ii) MMCS-DP and MMCS-MC (pressure sensitive materials), which was explained above. Indeed, in the cases when the edge of the yield surface is not reached, the MMCS-TR criterion mimics almost identically that of MMCS-vM (corresponding plots are hardly distinguishable from each other). For the pair: MMCS-DP and MMCS-MC, a worse resemblance is observed. Surprisingly, when the aforementioned condition is not met (the transition through the edge of yield surface occurs), the data for MMCS-TR and MMCS-MC match almost perfectly starting from the viscosity dominated regime ($\tilde p_0=0$) up to the moment when the solution kink takes place (see Fig.  \ref{Fig2COMP}). Clearly, when approaching the viscosity dominated regime ($\tilde p_0 \to \pi(1-\nu)$) all the results become identical. 

Moreover, from Fig. \ref{FigCOMP} and Fig. \ref{Fig2COMP} it can be seen that although the Poisson's ratio does affect the solution, generally in most cases the behaviour of $\theta_f$ and discrepancies between analysed criteria are very similar for each value of $\nu$.

Finally, we  recall the results from \cite{Perkowska_2017} obtained for the Maximum Circumferential Stress (MCS) criterion which does not account for the plastic deformation effect.  By comparing the graphs given in the cited paper with those presented here we conclude that the MCS criterion produces the angles of crack propagation very similar to those given by MMCS-DP and MMCS-MC over the entire range of analyzed loadings and parameters.

\begin{figure}[htb!]
\begin{center}
\includegraphics[scale=0.8]{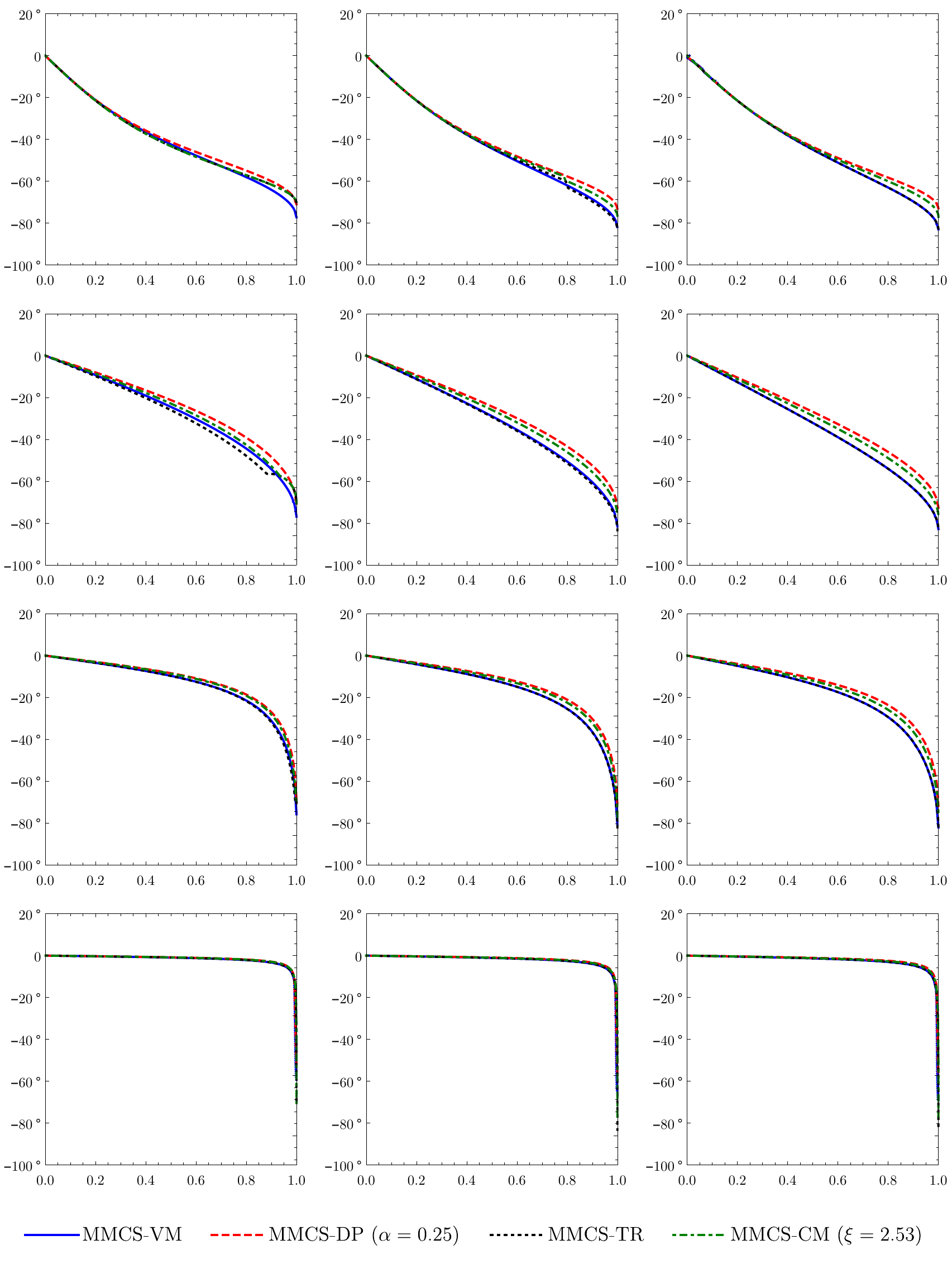}
{
\scriptsize
\put(-395,517){$\theta_f$}
\put(-395,394){$\theta_f$}
\put(-395,270){$\theta_f$}
\put(-395,146){$\theta_f$}
\put(-328,24){$\hat K_{II}$}
\put(-194,24){$\hat K_{II}$}
\put(-62,24){$\hat K_{II}$}
\put(-375,507){$\frac{\tilde p_0}{\pi(1-\nu)}=0$}
\put(-375,384){$\frac{\tilde p_0}{\pi(1-\nu)}=0.5$}
\put(-375,260){$\frac{\tilde p_0}{\pi(1-\nu)}=0.9$}
\put(-375,136){$\frac{\tilde p_0}{\pi(1-\nu)}=0.999$}
\put(-294,507){$\nu=0$}
\put(-168,507){$\nu=0.3$}
\put(-35,507){$\nu=0.5$}
}
\caption{Predicted propagation angle $\theta_f$ for various $\hat K_{II}$ and fixed $\tilde p_0$ and $\nu$. All analysed criteria are presented in each graph.}
\label{FigCOMP}
\end{center}
\end{figure}

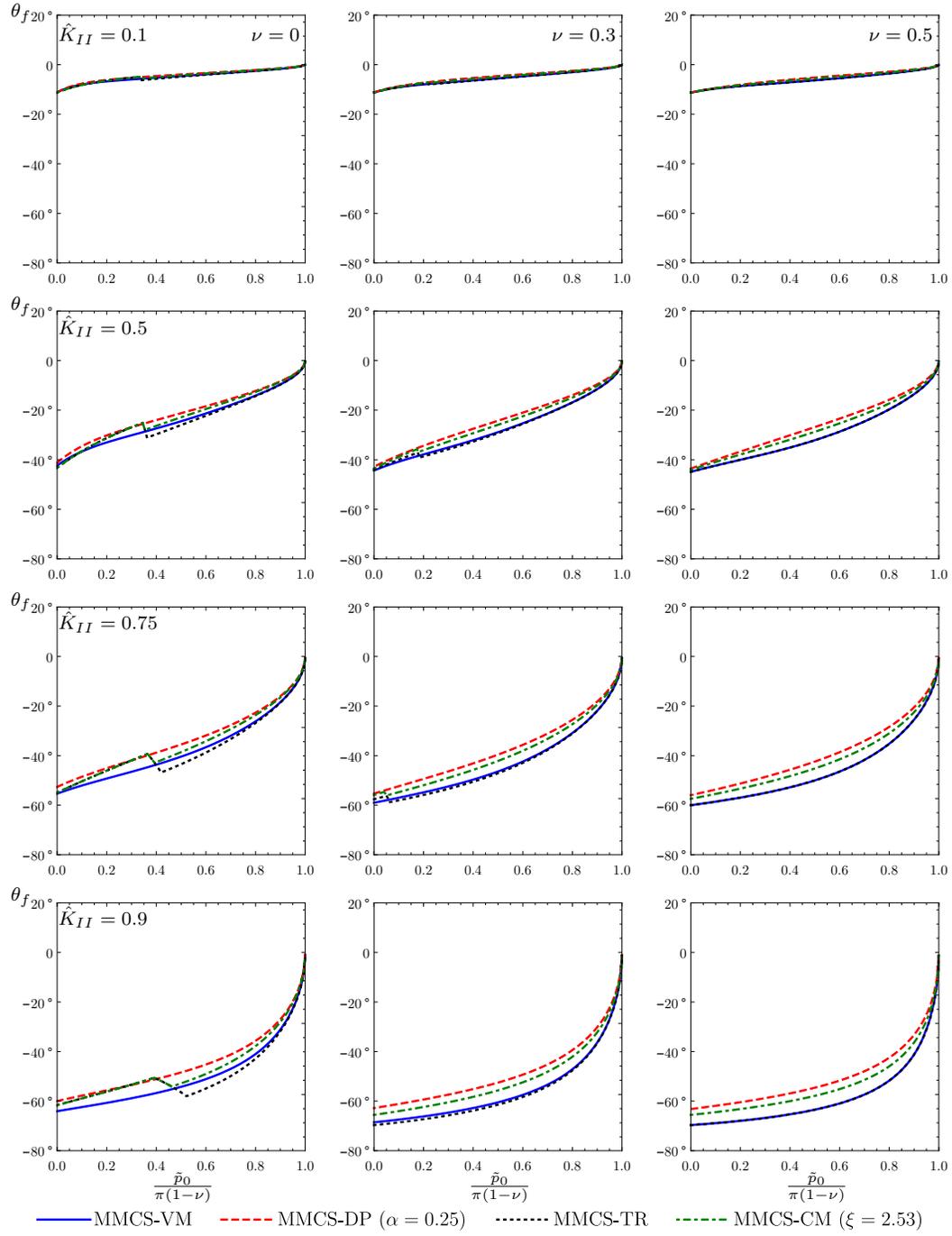
\begin{figure}[htb!]
\begin{center}
\includegraphics [scale=0.8]{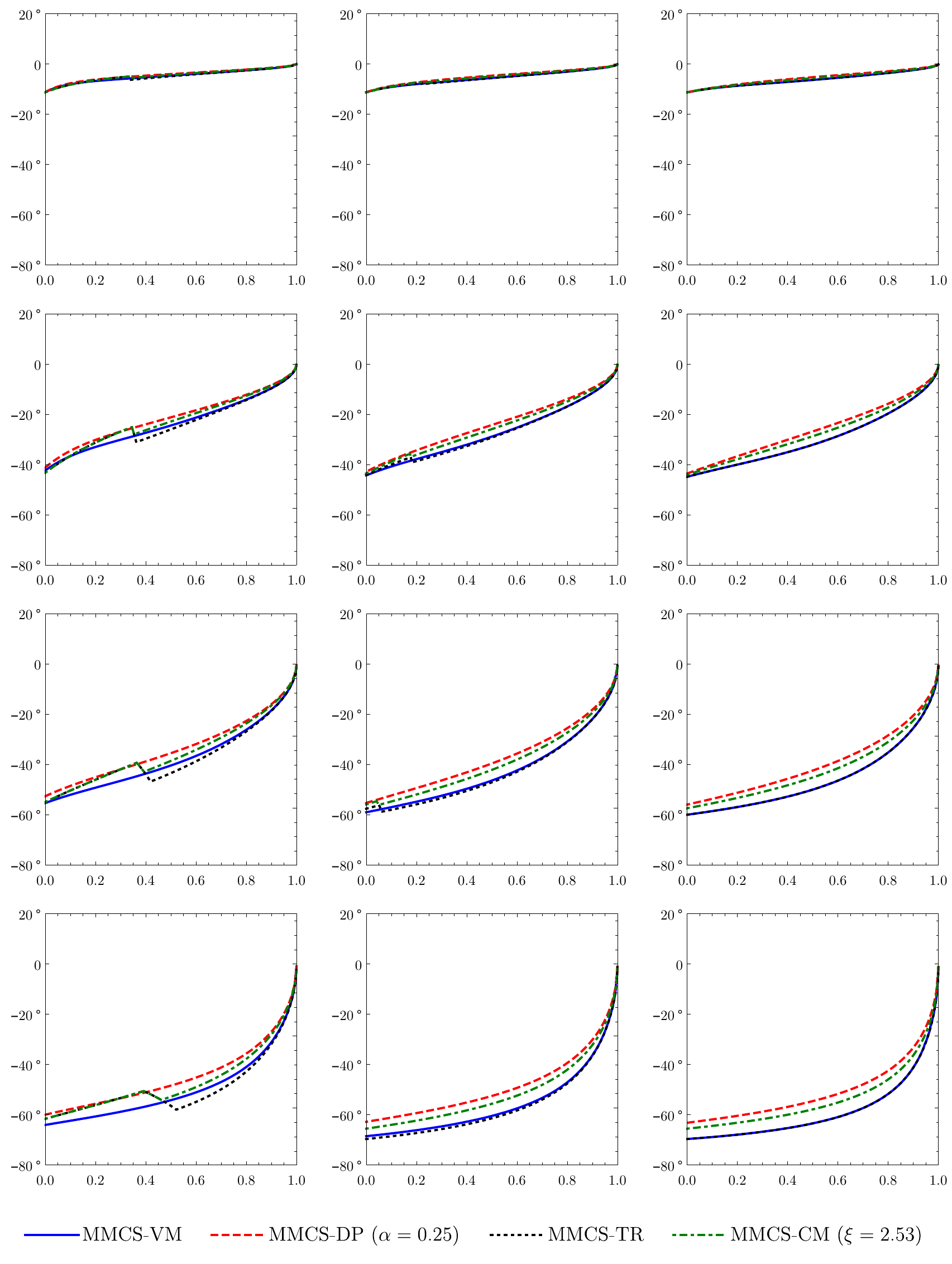}
{
\scriptsize
\put(-395,517){$\theta_f$}
\put(-395,394){$\theta_f$}
\put(-395,270){$\theta_f$}
\put(-395,146){$\theta_f$}
\put(-336,24){$\frac{\tilde p_0}{\pi(1-\nu)}$}
\put(-203,24){$\frac{\tilde p_0}{\pi(1-\nu)}$}
\put(-70,24){$\frac{\tilde p_0}{\pi(1-\nu)}$}
\put(-375,507){$\hat K_{II}=0.1$}
\put(-375,384){$\hat K_{II}=0.5$}
\put(-375,260){$\hat K_{II}=0.75$}
\put(-375,136){$\hat K_{II}=0.9$}
\put(-294,507){$\nu=0$}
\put(-168,507){$\nu=0.3$}
\put(-35,507){$\nu=0.5$}
}
\caption{Predicted propagation angle $\theta_f$ for various $\frac{\tilde p_0}{\pi(1-\nu)}$ and fixed $\hat K_{II}$ and $\nu$. All analysed criteria are presented in each graph.}
\label{Fig2COMP}
\end{center}
\end{figure}

\clearpage

\section{Conclusions}
\label{sec:conclusions}
In the paper a problem of redirection of a fluid driven crack was considered for a mixed mode loading that includes the classical Modes I-III and hydraulically induced tangential tractions on the fracture walls. The effect of plastic deformation in the near-tip zone was taken into account. Different criteria based on various yield conditions were employed to define the angle of fracture deflection.

The following conclusions can be drawn from the conducted analysis:
\begin{itemize}
\item{The component of loading related to the hydraulically induced shear stress has a substantial influence on the crack orientation. Its magnification under a fixed Mode II loading reduces the angle of fracture deflection. High sensitivity of the angle to the magnitude of loading is observed for a combination of severe antisymmetric shear (Mode II) and the so called viscosity dominated regime of fracture propagation. In such a case the actual angle of fracture deflection depends on the loading history.}
\item{Among the analyzed criteria it is the one based on the Maximum Dilatational Strain Energy Density (MDSED) that seems to be the least credible. It stems from the fact that in a certain range of loading parameters near the viscosity dominated regime it either produces no solution or yields unstable results. }
\item{Strong similarities between the results are observed for those criteria that use related yield conditions, i. e. for the pairs: i) MMCS-vM and MMCS-TR, ii) MMCS-DP and MMCS-MC (pressure sensitive materials).}
\item{The angle of crack propagation is always a smooth function of the loading components ($\tilde p_0$ and $\hat K_{II}$) only for those criteria which employ yield surfaces that are quadratics in the principal stresses space (MMCS-VM and MMCS-DP). For MMCS-TR and MMCS-MC, depending on the configuration of loading, the angle of fracture deflection can be only a continuous function (locally not smooth).}
\item{ The Maximum Circumferential Stress (MCS) criterion presented in \cite{Perkowska_2017} (which does not account for the plastic deformation effect) produces similar results to those obtained for  MMCS-DP ans MMCS-MC over the entire range of analyzed parameters. This suggests that, at least in some cases, it may be sufficient to base the analysis on the Linear Elastic Fracture Mechanics without a need to account for the plastic deformation.}
\item{Even though the investigated criteria produce very similar results it needs experimental verification to decide which one is the most credible for a predefined type of fractured material and loading components. Special attention should be devoted here to establish whether the applied solid material model has to be pressure sensitive or not.}
\end{itemize}

\vspace{5mm}

\noindent
{\bf Compliance with Ethical Standards}

\vspace{3mm}

\noindent
{\bf Funding:} This work was funded by European Regional Development Fund and the Republic of Cyprus through the Research Promotion Foundation (RESTART 2016 - 2020 PROGRAMMES, Excellence Hubs, Project EXCELLENCE/1216/0481) (PP, MW) and by the ERC Advanced Grant ”Instabilities and nonlocal multiscale modelling of materials” under number ERC-2013-ADG-340561-INSTABILITIES (AP, GM). The authors declare that these publicly funded arrangements have not created a conflict of interest.

\section*{Acknowledgments}
GM gratefully acknowledges support from a grant No. 14.581.21.0027 unique identifier: RFMEFI58117X0027 by Ministry of Education and Science of the Russian Federation and is thankful to the Royal Society for the Wolfson Research Merit Award. The authors are also thankful to Prof. Davide Bigoni and Dr Monika Perkowska for their useful comments and discussions.

\end{document}

%% file: definitions.tex
\newcommand{\beq}{\begin{equation}}
\newcommand{\eeq}{\end{equation}}

\newcommand{\beqar}{\begin{eqnarray}}
\newcommand{\eeqar}{\end{eqnarray}}
\newcommand{\bit}{\begin{itemize}}
\newcommand{\eit}{\end{itemize}}
\newcommand{\benum}{\begin{enumerate}}
\newcommand{\eenum}{\end{enumerate}}
\newcommand{\barr}{\begin{array}}
\newcommand{\earr}{\end{array}}
\def\ds{\displaystyle}

\def\XXint#1#2#3{{\setbox0=\hbox{$#1{#2#3}{\int}$}
   \vcenter{\hbox{$#2#3$}}\kern-.5\wd0}}

\def\b0{\mbox{\boldmath $0$}}

\newcommand{\bsigma}{\mbox{\boldmath $\sigma$}}

\def\f0{\ensuremath{\mathbb{O}}}

\DeclareMathOperator{\tr}{tr}
\newcommand{\dev}{\mathop{\mathrm{dev}}}



%% file: final_20_03_2019_arxiv.bbl
\begin{thebibliography}{99}

\bibitem{Bigoni_1993} Bigoni, D. and Radi, E. (1993) Mode I crack propagation in elastic-plastic pressure-sensitive materials. \textit{International Journal of Solids and Structures}, 30(7): 899-919
\bibitem{Bigoni_2004} Bigoni, D. and Piccolroaz, A. (2004) Yield criteria for quasibrittle and frictional materials. \textit{International Journal of Solids and Structures}, 41: 2855-2878.
\bibitem{Camas_2011} Camas D., Hiraldo I., Lopez-Crespo P., Gonzalez-Herrera A. (2011) Numerical and experimental study of mixed-mode cracks in non-uniform stress field. \textit{Procedia Engineering}, 10: 1691-1696.

\bibitem{cherny_2016} Cherny S., Lapin V.,  Esipov D., Kuranakov D.,  Avdyushenko A., Lyutov A.,  Karnakov P. (2016) Simulating fully 3D non-planar evolution of hydraulic fractures. \textit{International Journal of Fracture}, 201: 181-211.

\bibitem{cherny_2017} Cherny S., Esipov D., Kuranakov D., Lapin V., Chirkov D., Astrakova A. (2017) Prediction of fracture initiation zones on the surface of three-dimensional structure using the surface curvature. \textit{Engineering Fracture Mechanics},  172: 196-214.

\bibitem{van_Dam_1999}
van Dam D.B and de Pater C.J. (1999). Roughness of Hydraulic Fractures: The Importance of In-Situ Stress and Tip Processes, \textit{SPE}, 56596.

\bibitem{van_Dam_2002}
van Dam D.B, Papanastasiou P, De Pater C.J. (2002) Impact of rock plasticity on hydraulic fracture propagation and closure, \textit{J. SPE Production \& Facilities}, 17 (3): 149-159.

\bibitem{erdogan_1963} Erdogan F., Sih G.C. (1963) On the crack extension in plates under plane loading and transverse shear. \textit{Journal of Basic Engineering}, 85(4): 519-525.

\bibitem{Economides_2000} Economides M., Nolte K. (2000) Reservoir Stimulation, 3rd edn. Wiley, Chichester.

\bibitem{Hallback_1994}  Hallback N., Nillson F. (1994) Mixed-mode I/II fracture behavior of an aluminum alloy. \textit{Journal of the Mechanics and Physics of Solids }, 42(9): 1345-1374.

\bibitem{lazarus_2008} Lazarus V., Buchholz F.-G., Fulland M., Wiebesiek J. (2008) Comparison of predictions by mode II or mode III criteria on crack front twisting in three or four point bending experiments. \textit{International Journal of Fracture}, 153(2): 141-151.

\bibitem{Papanastasiou_1993} Papanastasiou P., Thiercelin M. (1993) Influence of inelastic rock behaviour in hydraulic fracturing. \textit{International Journal of Rock Mechanics and Mining Sciences \& Geomechanics Abstracts}, 30: 1241-1247.

\bibitem{Papanastasiou_1997} Papanastasiou P. (1997) The influence of plasticity in hydraulic fracturing. \textit{International Journal of Fracture}, 84: 61-97.

\bibitem{Papanastasiou_1999} Papanastasiou P. (1999) The effective fracture toughness in hydraulic fracturing. \textit{International Journal of Fracture}, 96: 127-147.

\bibitem{Papanastasiou_2017} Papanastasiou P., Durban D. (2018) The influence of normal and shear stress loading on hydraulic fracture-tip singular plastic fields. \textit{Rock Mechanics and Rock Engineering}, 51(10): 3191–3203.

\bibitem{Perkowska_2017} Perkowska M., Piccolroaz A., Wrobel M., Mishuris G. (2017) Redirection of a crack driven by viscous fluid, \textit{ International Journal of Engineering Science }, 121: 182-193.

\bibitem{Sarris_2011} Sarris E., Papanastasiou P. (2011) The influence of the cohesive process zone in hydraulic fracturing, \textit{International Journal of Fracture}, 167: 33-45.

\bibitem{Sarris_2013} Sarris E., Papanastasiou P. (2013) Numerical Modelling of Fluid-Driven Fractures in Cohesive Poro-elastoplastic Continuum, \textit{International Journal for Numerical and Analytical Methods in Geomechanics}, 37(12): 1822-1846.

\bibitem{sih_1974} Sih G.C. (1974) Strain-energy-density factor applied to mixed mode crack problems. \textit{International Journal of Fracture}, 10(3): 305-321.

\bibitem{theocaris_T_1982} Theocaris  P.S.,  Andrianopoulos  N.P. (1982) The T-Criterion applied to ductile fracture.  \textit{International Journal of Fracture}, 20: R125-R130.

\bibitem{Wang_2015} Wang H. (2015) Numerical Modeling of Non-Planar Hydraulic Fracture Propagation in Brittle and Ductile Rocks using XFEM with Cohesive Zone Method, \textit{Journal of Petroleum Science and Engineering}, 135: 127-140.

\bibitem{Weinberger_1994} Weinberger R., Reches 2.. Eidelman A., Scott T.S. (1994) Tensile properties of rocks in four-point beam tests under confining pressure, in Proceedings First North American Rock Mechanics Symposium, Austin, Texas, eds Nelson, P. $\&$ Laubach, S.E., pp. 435-442

\bibitem{wrobel_2015} Wrobel M., Mishuris G. (2015) Hydraulic fracture revisited: Particle velocity based simulation. \textit{International Journal of Engineering Science}, 94: 23-58.

\bibitem{wrobel_2017} Wrobel M., Mishuris G., Piccolroaz A. (2017) Energy Release Rate in hydraulic fracture: can we neglect an impact of the hydraulically induced shear stress? \textit{International Journal of Engineering Science}, 111: 28-51.

\bibitem{wrobel_2018} Wrobel M., Mishuris G., Piccolroaz A. (2018) On the impact of tangential traction on the crack surfaces induced by fluid in hydraulic fracture: Response to the letter of A.M. Linkov. Int. J. Eng. Sci. (2018) 127, 217–219. \textit{International Journal of Engineering Science}, 127: 220-224.

\bibitem{yehia_Y_1991} Yehia N.A.B. (1991) Distortional strain energy density criterion: the Y-Criterion. \textit{Engineering Fracture Mechanics},  39(3): 477-485.


\end{thebibliography}
